# Candidate for a Fractional Topological Insulator in Twisted $MoTe_2$

Yiping Wang[1,2,*], Gillian E. Minarik[1,*], Weijie Li[3], Yves Kwan[4], Shuai Yuan[3], Eric Anderson[3], Chaowei Hu[5], Julian Ingham[6], Jeongheon Choe[1], Takashi Taniguchi[7], Kenji Watanabe[8], Xavier Roy[1], Jiun-Haw Chu[5], Raquel Queiroz[6], James C. Hone[2], N. Regnault[9,10,11], Xiaodong Xu[3,5], and Xiaoyang Zhu[1,†]

[1] Department of Chemistry, Columbia University, New York, NY 10027, USA

[2] Department of Mechanical Engineering, Columbia University, New York, NY 10027, USA

[3] Department of Physics, University of Washington, Seattle, WA 98195, USA

[4] Department of Physics, University of Texas at Dallas, Richardson, TX 75080, USA

[5] Department of Materials Science and Engineering, University of Washington, Seattle, WA 98195, USA

[6] Department of Physics, Columbia University, New York, NY 10027, USA

[7] Research Center for Materials Nanoarchitectonics, National Institute for Materials Science, 1-1 Namiki, Tsukuba 305-0044, Japan

[8] Research Center for Electronic and Optical Materials, National Institute for Materials Science, 1-1 Namiki, Tsukuba 305-0044, Japan

[9] Center for Computational Quantum Physics, Flatiron Institute, 162 5th Avenue, New York, NY 10010, USA

[10] Department of Physics, Princeton University, Princeton, New Jersey 08544, USA

[11] Laboratoire de Physique de l'Ecole Normale Supérieure, PSL University, CNRS, Sorbonne Universités, Université Paris Diderot, Sorbonne Paris Cité, Paris, France

**The interplay among electronic correlation, topology, and time-reversal-symmetry (TRS) often leads to exotic quantum states of matter, as highlighted by the discoveries of fractional Chern insulators (FCIs) in twisted bilayer $MoTe_2$ ($tMoTe_2$). Among the FCIs in $tMoTe_2$, the most robust is at a hole filling factor of $\nu = -\frac{2}{3}$ per moiré unit cell. Here, employing pump-probe circular dichroism (CD) measurement on $tMoTe_2$ at twist angles θ = 3.9° and 3.7°, we show that a correlated state at $\nu = -\frac{4}{3}$ exhibits an unusual Ising antiferromagnet behavior. The $\nu = -\frac{4}{3}$ state with no net magnetization undergoes first order phase transitions at**

[*] These authors contributed equally.
[†] To whom correspondence should be addressed. Email: xyzhu@columbia.edu

**extremely low magnetic fields of $|\mu_0 H|$ ~ 2-6 mT to partially valley polarized (PVP) states. This behavior is notably absent for all other correlated states in $tMoTe_2$ and also disappears for $\nu = -\frac{4}{3}$ at higher or lower twist angles (θ = 4.0° or 3.3°). The observed magnetic signature is consistent with a theoretically proposed fractional topological insulator (FTI), consisting of two copies of $\nu_\pm = -\frac{2}{3}$ FCIs with opposite chirality in the $K^\pm$ valleys. The experimental results are supported by interacting continuum model calculations that reveal the extreme closeness in energy (ΔE < 1 meV) between the putative FTI ($\nu_\pm = -\frac{2}{3}$ and $\nu_\mp = -\frac{2}{3}$) and PVP states ($\nu_\pm = -1$ and $\nu_\mp = -\frac{1}{3}$). Our findings present a candidate FTI with TRS and call for advanced transport and imaging measurements to establish the quantized helical edge modes.**

The discoveries of FCIs and the associated fractional quantum anomalous Hall (FQAH) effects, in twisted $MoTe_2$ bilayers [1–4] and in rhombohedral multilayer graphene aligned with hexagonal boron nitride (hBN) [5], represent milestones in the search for topological quantum matters. They are experimental realizations of the much-celebrated lattice models [6–8] for the fractional quantum Hall (FQH) effect in the absence of external magnetic field [9–11]. In these moiré systems, TRS is spontaneously broken and valley degeneracy lifted in partially filled flat Chern bands, a combined result of topology and strong correlation. For the partial filling of the first Chern band in $tMoTe_2$, spontaneous TRS breaking has been found to occur in a finite hole doping window of $-1.3 < \nu < -0.4$ (per moiré unit cell). Interestingly, at the boundary between non-magnetic and magnetic regions, transient optical sensing [12] and more recent transport measurements [13,14] revealed a correlated state at $\nu = -\frac{4}{3}$, which is twice the filling factor of the most robust $\nu = -\frac{2}{3}$ FCI in $tMoTe_2$ [1–4]. While two copies of the $\nu = -\frac{2}{3}$ FCIs in opposite $K$ valleys have been intuitively assumed to form an FTI with TRS [15], computational analysis revealed the fragile nature of the $\nu = -\frac{4}{3}$ FTI in $tMoTe_2$ and suggested potential mechanisms for its stabilization [16].

We investigate four dual-gated $MoTe_2$ bilayer devices (D1-4) with twist angles θ = 3.90±0.06 ° (D1), 3.70±0.07 ° (D2), 4.00±0.08 ° (D3), and 3.30±0.07 ° (D4). Devices at 3.6° ≤ θ ≤ 4.0° are known to host FCI states, particularly the most robust at $\nu = -\frac{2}{3}$ [1–4,13,17]. Device fabrication and characterization are detailed in the appendix (Methods, Extended Data Fig. 1), and in a previous report [13]. We have developed a pump-probe time-domain approach, Fig. 1a, to probe moiré quantum phases [18,19] and demonstrated its sensitivity in detecting a large number of correlated states, including incipient states, in $tMoTe_2$ [12]. The extraordinary sensitivity lies in its background free nature when the pump pulse selectively targets the correlation gap or pseudo gap (Methods). This is essentially a form of modulation spectroscopy primarily probing the ground state – after a few picoseconds, the pump's disruption of correlation in the moiré quantum phase provides the only contrast in optical response, as isolated by the differential reflectance ΔR/R, where ΔR is pump-induced spectral change and R is reflectance spectrum before pump [12]. The results presented in the following are independent of pump light polarization, consistent with non-selective excitation across the correlation gaps, and with the spectral modulation amplitude increasing with pump excitation pulse energy density (7-42 $\mu J \cdot cm^{-2}$) (Extended Data Fig. 2).

Here we combine the extraordinary sensitivity of transient reflectance with circular dichroism (CD) to quantify the spin/valley polarization based on optical selection rules for exciton-Fermi polarons [20,21]. An attractive polaron (AP) consists of an exciton in the $K^-$ ($K^+$) valley bound to carriers in the opposite $K^+$ ($K^-$) valley, Fig. 1b. Due to spin-valley locking, the transition is selectively probed by $\sigma^-$ ($\sigma^+$) circularly polarized light and this approach has been applied to the quantification of spin/valley polarization in the region of spontaneous TRS breaking, including Chern insulators, in $tMoTe_2$ [22,23]. At $-1.3 < \nu < -0.4$ , the AP transition emerges for only one circular polarization because the hole population resides entirely in one valley ($-1 \leq \nu < -0.4$) or is highly polarized towards one valley ($-1.3 < \nu < -1$) [1–4,24,25].

We present transient reflectance spectra as a function of $\nu$ with $\sigma^+$ (Fig. 1c) and $\sigma^-$ (Fig. 1d) circularly polarized probe light on device D1 (θ = 3.9°), without external magnetic field ($\mu_0 H$ = 0 mT). While we resolve correlated states with either $\sigma^+$ or $\sigma^-$ polarization, the spectral amplitudes are different, as expected from the different transition strengths in the presence of spin/valley polarization. In the doping region $-1.3 < \nu < -0.4$, the AP resonance (~1.2-1.3 eV) from $\sigma^+$ (Fig. 1c) is stronger than that from $\sigma^-$ (Fig. 1d) because holes are polarized in the $K^-$ valley (negative $\mu_0 H$ field training). We show in Fig. 1e the difference spectra, $|\sigma^+\rangle - |\sigma^-\rangle$, which we take as a proxy to CD. The CD spectra show non-zero intensity only in the region $-1.3 < \nu < -0.4$, where spontaneous spin/valley polarization is known to occur [1–4,24,25]. While the states are not well-resolved for either $\sigma^+$ or $\sigma^-$ polarizations, they are more obvious in the CD spectra, clearly showing the integer and fractional Chern insulators at $\nu =$ -1 and $-\frac{2}{3}$, respectively. Interestingly, the transient CD spectra also reveal states slight above or below ν ~ -1; the nature of these states is not known and deserves future investigations. In the following, we focus on the $\nu = -\frac{4}{3}$ and $-\frac{3}{2}$ states that have been detected in transient optical sensing [12] and confirmed in transport [13,14,25]. The $\nu = -\frac{4}{3}$ and $-\frac{3}{2}$ states show distinct signatures in transient spectral sensing.

For $\nu = -\frac{3}{2}$, which is believed to be a charge ordered states such as a Wigner crystal or a charge density waves (CDW) [13,25], we observe strong AP resonance with either $\sigma^+$ or $\sigma^-$ polarization, Fig. 1c and Fig. 1d. Similar behaviors are observed for other charge ordered states, such as $\nu = -\frac{1}{3}$ [13,25] and a large number of states on the electron doping side [12] (see Extended Data Fig. 3 for broader $\nu$ ranges). These states are non-magnetic, evident by the vanishing CD signal. Interestingly, previous optical sensing in $tMoTe_2$ revealed negligible exciton or AP oscillator strength for hole doping at $\nu < -1.2$ [22,23]. The much-enhanced AP oscillator strength observed for $\nu = -\frac{3}{2}$ may be a result of charge localization, which redistributes the charge population across broad momentum spaces, leading to reduced phase-space filling effect in the specific $K^+$ or $K^-$ valley probed by $\sigma^+$ or $\sigma^-$light. The same argument may explain the high AP oscillator strengths for other charge ordered states on the electron doping side at $\nu \geq 1$.

In contrast to $\nu = -\frac{3}{2}$, the spectroscopic response of the nearby $\nu = -\frac{4}{3}$ state is distinctly different. For $\sigma^+$ or $\sigma^-$ probe, both exciton and AP transitions at $\nu = -\frac{4}{3}$ are much weaker than those at either higher or lower hole fillings. The $\nu = -\frac{4}{3}$ state also shows zero intensity in CD spectra at zero magnetic field, Fig. 1e, confirming the absence of spontaneous ferromagnetism [1–4,24,25]. When we add a small out-of-plane magnetic field of $\mu_0 H$ = -6 mT, Figs. 1f-h, a new signature emerges: we observe a strong AP resonance at $\nu = -\frac{4}{3}$ for $\sigma^+$ polarization (Fig. 1f), but not for $\sigma^-$ polarization (Fig. 1g). This reveals that hole population is polarized to the $K^-$ valley upon the application of $\mu_0 H$ = -6 mT. Fig. 1h now shows that the CD signal of the $\nu = -\frac{4}{3}$ state is comparable to those in the $-1.3 < \nu < -0.4$ region. Thus, a small magnetic field switches the $\nu = -\frac{4}{3}$ state from one with no magnetization to a spin/valley polarized state. In stark contrast, the switching behavior is absent for all other correlated states, including the nearby $\nu = -\frac{3}{2}$.

The magnetic behavior of the $\nu = -\frac{4}{3}$ state is established in $\mu_0 H$-field scans. We present transient reflectance spectra as a function of $\mu_0 H$ with $\sigma^+$ and $\sigma^-$ polarized probes for devices D1 (θ = 3.9°, Figs. 2a-i) and D2 (θ = 3.7°, Figs. 2j-r) at the fixed filling factor of $\nu = -\frac{4}{3}$. We note that Fig 2. zooms in on the spectral window covering the AP transition, the only branch exhibiting nonzero CD at $\nu = -\frac{4}{3}$ with finite magnetic fields. We first focus on D1. Starting from $\mu_0 H$ = -12 mT and moving to positive values (Fig. 2a), we observe strong AP resonance from $\sigma^+$ polarization, indicating holes predominantly in the polarized $K^-$ valley. As the $\mu_0 H$ field reaches a critical value of ~ $-4$ mT, we observe a sudden decrease in the AP resonance intensity by a factor of ~30 ×, suggesting a first-order phase transition to a state where this transition is much weakened. While the $\sigma^+$-AP transition remains weak as $\mu_0 H$ is further increased to positive values, the $\sigma^-$-AP transition is abruptly increased for $\mu_0 H \geq 5$ mT, Fig. 2b, indicating holes predominantly polarized in the $K^+$ valley. At $-4$ mT < $\mu_0 H$ < 5 mT, both $\sigma^+$ and $\sigma^-$ AP transitions are weak, suggesting that $K^+$ and $K^-$ valleys are equally populated. The presence of the three states is most obvious in the CD spectra $|\sigma^+\rangle - |\sigma^-\rangle$, Fig. 2c, which reveals a $K^-$ polarized state on the negative $\mu_0 H$ side, an intermediate state with equal $K^-$ and $K^+$ occupations, and a $K^+$ polarized state on the positive $\mu_0 H$ side. These three states are also observed as the direction of the $\mu_0 H$ field sweep is reversed, Figs. 2d-f, with clear hysteresis expected from magnetic transitions.

We extract the changes to AP oscillator strength, $\Delta f_{AP}^+$ and $\Delta f_{AP}^-$ for $\sigma^+$ and $\sigma^-$ polarized light from modeling the 2D optical susceptibility of the vdW stack using Fresnel's equations (see Methods). We show $\Delta f_{AP}^+$ (Fig. 2g) and $\Delta f_{AP}^-$ (Fig. 2h) and the corresponding valley/spin polarizations $< S_z > = \Delta f_{AP}^+ - \Delta f_{AP}^- / (\Delta f_{AP}^+ + \Delta f_{AP}^-)$ (Fig. 2i). While the magnetic field sweep for $\sigma^+$ (Fig. 2g) or $\sigma^-$ (Fig. 2h) selectively probes hole polarization in the $K^-$ or $K^+$ valley, $< S_z >$ reveals the transitions from $K^-$ polarized, $K^-$ and $K^+$ equally populated, and $K^+$ polarized states. We note that due to vanishing signal between $-4$ mT < $\mu_0 H$ < 5 mT, we are unable to extract meaningful $\Delta f_{AP}^{\pm}$ values for the non-spin-polarized sectors in D1 and D2. Consequently, $\Delta f_{AP}^{\pm}$ is zero in Figs. 2g-h, where AP features are absent to indicate suppression of the optical transition. The transition near the critical field is hysteretic and sharp, resembling an Ising AFM state under the application of magnetic field along the easy axis (surface normal). Note that we use the term, Ising AFM, to describe magnetic behavior in momentum space, not the conventional meaning of AFM in real space. In each $\sigma^+$ or $\sigma^-$ valley-selective probe, the hysteresis loop is similar to that of a ferromagnetic response. It is the simultaneous presence of ferromagnetic responses in both $K^-$ and $K^+$ valleys which gives rise to the Ising AFM behavior referred to here.

The observed magnetic behavior is confirmed in $\mu_0 H$-field scans on device D2 ($\theta = 3.7^o$), Figs. 2j-r. In this device, the $\nu = -\frac{4}{3}$ state again exhibits zero magnetization in a finite $\mu_0 H$ field window around 0 mT, switching to positive/negative magnetization outside the window. In each scan direction (Fig. 2j,k or Fig. 2m,n), we observe the transitions from a $K^-$ polarized state to an intermediate state with equal $K^-$ and $K^+$ occupation, and a $K^+$ polarized state on the positive $\mu_0 H$ side. For either $\sigma^+$ or $\sigma^-$ probe, we observe hysteresis in the two $\mu_0 H$-field scan directions, as shown by AP oscillator strength, $\Delta f_{AP}^{\pm}$, in Fig. 2p, 2q. The magnetic behavior is summarized in the valley spin polarizations $< S_z >$, Fig. 2r. Except for the sharp transitions, the ΔR/R signal remains flat, indicating zero magnetic susceptibility.

The observed Ising AFM behavior is qualitatively robust, as shown by $\mu_0 H$-field scans of $< \mathrm{S_z} >$ in spatially resolved measurements on sample D1, Extended Data Fig. 4. However, from sampling spot to sampling spot, there are quantitative differences in the range and degree of asymmetry in the forward and backward $\mu_0 H$-field scans, which is an indication of sensitivity to disorder, e.g., small variations in twist angle and strain field. Note that the measurements in Extended Data Fig. 4 was made after a few cycles of warm-up (to room temperature) and cool-down in the cryostat – a procedure known from our experience and anecdotes in the field to result in changes in moiré properties, likely due to cycles of expansion and contraction strains. Interestingly, the AFM behavior in Extended Data Fig. 4 from thermally cycled sample D1 resembles those in Fig. 2j-r from sample D2, again pointing to the sensitivity of the state to local variation in moiré structure.

In all measurements showing AFM behavior, the observed flatness of the zero-magnetization region with applied $\mu_0 H$ field, i.e., zero magnetic susceptibility, suggests the $\nu = -\frac{4}{3}$ state is gapped in each valley, consistent with the resistive behavior observed in transport measurements [13,14]. For comparison, CD becomes measurable around $\mu_0 H = 0$ mT when the AFM-like behavior disappears at a higher temperature (device D1, θ = 3.9°, T = 6 K) or smaller twist angle (device D4, θ = 3.3°, T = 1.6 K); see below. As shown in Extended Data Fig. 5, the magnetization of the spin/valley polarized states in Fig. 2i at $|\mu_0 H| \geq 5$ mT remains constant as the external magnetic field is increased by three orders of magnitude to $|\mu_0 H|$ = 6T, confirming that further hole population transfers are not possible. The saturation behavior suggests that these states are PVPs with $\nu^+$ $(\nu^-)$ = -1 in the $K^+$ $(K^-)$ valley and $\nu^-$ $(\nu^+)$ $= -\frac{1}{3}$ in the $K^-$ $(K^+)$ valley.

In both D1 and D2, the magnetic signature persists at all measured time delays (up to Δt = 1.2 ns) with identical hysteresis at exactly the same $\mu_0 H$ fields, varying only in ΔR/R intensity (Extended Data Fig 6). This confirms that the observed Ising AFM behavior at $\nu = -\frac{4}{3}$ is repeatable and intrinsic to the doped moiré devices, rather than a transient light-induced effect. Moreover, our transient experiment disrupts and restores the ground state for each pump-probe pulse pair at 400 kHz. Reproducibility from pulse-to-pulse suggests that the hysteretic magnetic phase transitions are unlikely a result of a state consisting of equal populations of magnetized domains of opposite signs, as such a hypothetical domain pattern is not expected to be exactly reproduced after each pump-probe laser pulse pair. In light of recent work [23,26,27], we also rule out optical switching of the Chern number by using the optical power four orders of magnitude below the switching threshold (Methods). Thus, our transient modulation CD spectroscopy probes the magnetic behavior of the ground state. Supporting this conclusion, we confirm that the Ising AFM behavior is also observed in static CD measurement (Extended Data Fig. 7), albeit with order(s)-of-magnitude lower signal-to-noise ratios than those in transient CD spectroscopy.

We carry out temperature (*T*) dependent measurements to establish thermal stability of the $\nu = -\frac{4}{3}$ state for device D1 (θ = 3.9°). Fig. 3a shows transient reflectance spectra for $\sigma^+ - \sigma^-$ probe with forward-sweeping magnetic field at the four indicated temperatures, T = 1.6, 4, 6, and 10 K; the corresponding spectra for the $\sigma^+ - \sigma^-$ probe with reverse-sweeping magnetic field can be found in Extended Data Fig. 8. We also show the extracted AP oscillator strength, $\Delta f^{\pm}_{AP}$, in Fig. 3b. While the AFM behavior is robust at T = 1.6 K, it becomes unstable from spin fluctuations at T = 4 K and vanishes at T = 6 K and 10 K. Thus, the critical temperature for the AFM state at $\nu = -\frac{4}{3}$ is $T_c \sim 4$ K. We also determine the stability of the observed state with vertical displacement field and find the critical field for destroying the state to be of the order of $E_c \sim 0.1$V/nm (Extended

Data Fig. 8). Both $T_c$ and $E_c$ for the AFM-like state are similar to those for the stability of the $\nu = -\frac{2}{3}$ FCI at similar twist angles [1–4,24,25].

The AFM magnetic behavior at $\nu = -\frac{4}{3}$ exists at θ = 3.7° or 3.9° but disappears in $\mu_0H$- scans at lower or higher twist angles, θ = 3.3° (D3) and θ = 4.0° (D4). At θ = 4.0°, Fig. 3c, the magnetic field scan reveals a ferromagnetic state, i.e., the boundary of spontaneous TRS breaking nominally observed for $-1.3 < \nu < -0.4$ is extended to $\nu = -\frac{4}{3}$, a property which has eluded detection in static optical sensing. Thus, the AFM $\nu = -\frac{4}{3}$ state emerges at the boundaries of competing orders. At θ = 3.3°, Fig. 3c, there is no evidence for magnetic order for external magnetic fields as high as 6 T (Extended Data Fig. 9). Note that the magnetic field scans in AP oscillator strength for D1 at 6 K (Fig. 3b3) and D4 at 1.6 K (Fig. 3c4) show weak paramagnetic behavior, in contrast to the flat zero CD around zero field in Fig. 2.

The Ising AFM behavior observed here at $\nu = -\frac{4}{3}$ is unexpected from the known ferromagnetism in the broad doping region of $-1.3 < \nu < -0.4$. The latter is believed to result from a Stoner mechanism due to the dominant direct exchange interaction [28]. The Stoner criterion for ferromagnetism, $I \cdot DOS(E_F) > 1$ [$I$ is exchange energy and $DOS(E_F)$ the density of states], is satisfied because of the high density of states and because of the significant overlap of wavefunctions between the two sites in each moiré unit cell and between neighboring unit cells. Ising AFM has not been reported before in the large body of literature on moiré quantum phases. The most abundant charge ordered states in moiré superlattices are paramagnetic [29], with tendencies of in-plane ferromagnetic or antiferromagnetic interactions due to kinematic [30] or frustrated [31] mechanisms, respectively. In twisted $WSe_2$ bilayers, in-plane AFM orders have been predicted theoretically [32,33] and observed at integer fillings under displacement field [29]. An in-plane AFM ordered state should show gradual canting under an out-of-plane magnetic field, not the abrupt first order phase transitions observed here. In $tMoTe_2$, the $\nu = -2$ state is theoretically predicted to be of in-plane AFM orders [34] while experimentally assigned to a quantum spin Hall (QSH) state [35]. Even for a QSH state, similar phase transitions under weak magnetic field are not expected because both $K^+$ and $K^-$ valleys are fully occupied in the first moiré band. This is confirmed in our experiment: the $\nu = -2$ state in D1 (3.9°) exhibits vanishing CD within the range ± 50 mT (Extended Data Fig. 10). Only when Zeeman-field-induced band-crossing occurs between the first Chern band in one valley ($K^+$ or $K^-$) and the second Chern band in the opposite can a magnetic switching occur for the QSH at $\nu = -2$, at $\mu_0 H$ field as high as ~14 T and ~7 T in $tMoTe_2$ at θ = 3.8° and 2.6°, respectively [25].

We now consider the FTI interpretation of the $\nu = -\frac{4}{3}$ state in $tMoTe_2$ [16]. The FTI, which is considered equivalent to the fractional quantum spin Hall (FQSH) effect in 2D, has been subject of considerable theoretical interest [36–39], yet its experimental realization has remained elusive. The most attractive proposal to obtaining FTIs comes from the interacting lattice model where two copies of FCIs with opposite chirality combine to form an FTI with necessary time-reversal symmetry [40–42]. In view of discoveries of FCIs in topological moiré flat-bands [1–5,15], one naturally asks the question on whether FTIs also exist in these systems. Kang et al. [35] assigned a $\nu = -3$ state in $tMoTe_2$ at a relatively small twist angle of θ = 2.1° to a fractional QSH state in the second moiré band and this interpretation has attracted considerable theoretical interest [16,43–50]. However, this state is ferromagnetic, breaks TRS, and shows finite $R_{xy}$ in Hall measurement at zero magnetic field [25,51,52]. In particular, Li et al. carried out a detailed mapping of magnetic phases and showed universal ferromagnetism around $\nu = -3$ in $tMoTe_2$ in a broad twist angle range of θ = 2.1° – 3.7° [52]. More importantly, these authors detected no topological gap in the TRS breaking state at $\nu = -3$.

In contrast to the earlier report at $\nu = -3$ [35], an FTI at $\nu = -\frac{4}{3}$ preserving TRS can form from two known FCIs at $\nu = -\frac{2}{3}$ with opposite chirality in $tMoTe_2$ [16,53], as has long been proposed in the interacting lattice model [40–42]. Such an FTI is predicted to be stabilized with screening of the Coulomb potential, but only in a small twist angle range around θ ~ 3.7° [16]; the latter is consistent with our experimental observations. The presence of two opposite chirality copies of the $\nu = -\frac{2}{3}$ FCIs preserves TRS, which can be broken by an out-of-plane magnetic field to switch to a PVP state with $\nu = -1$ and $\nu = -\frac{1}{3}$ in the two $K^+/K^-$ valleys. This leads to AFM→FM spin-flip magnetic transitions, with characteristic hysteresis under the application of a magnetic field, as is observed here for θ = 3.9° and 3.7° (Figs. 2 and Extended Data Fig. 4). In addition to the arithmetic, $-\frac{4}{3} = 2 \times \left(-\frac{2}{3}\right)$, it is also not surprising that the putative FTI exists at $\nu = -\frac{4}{3}$, which is at the phase boundary between nonmagnetic and magnetic regions.

To further support the FTI interpretations, we carry out ED calculations at $\nu = -\frac{4}{3}$ on the continuum model of $tMoTe_2$ (Fig. 4a, see Method section 8.1) for θ = 3.4°-4.4°. In Fig. 4b, we plot the energy difference (ΔE) between the ground states in the PVP ($\nu^+$ = -1 and $\nu^-$ = -1/3) and non-polarized ($\nu^+ = -\frac{2}{3}$ and $\nu^- = -\frac{2}{3}$) sectors on a 12-site lattice. The computational complexity requires projection to the lowest valence band, but we still incorporate crucial band-mixing effects [54–56] by allowing a restricted number of holes $N^1_{max}$ to occupy the second valence band (labelled band1 in Fig. 4b). While the PVP sector has lower energy in the absence of band-mixing, increasing $N^1_{max}$ can lead to a transition to the non-polarized sector. The energy difference remains small (fraction of meV), signaling the extremely close competition between the valley sectors, as is reflected in the small critical magnetic fields in switching between the states in Fig. 2. Calculations on other system sizes (see Method section 8.2) reveal similar qualitative behaviors, though finite-size effects preclude a more quantitative analysis. We also find an analogous depolarizing effect of band-mixing at $\nu = -\frac{3}{2}$ (Method section 8.4).

In order to access larger system sizes and evaluate the relevance of different many-body phases, we implement 36-site variational ED calculations that are mostly restricted to the lowest valence band (Method section 8.3). It has been theoretically argued that mixing with higher valence bands can weaken the interaction between the valleys and induce depolarization of the system as in Fig. 4b. Here, we introduce a phenomenological tuning parameter that simply multiplies the intervalley coupling by a factor $\lambda \leq 1$. Fig. 4c shows that the ground state lies in the fully valley-polarized sector ($\nu_+ = -\frac{4}{3}$ and $\nu_- = 0$) for λ=1, and sequentially transitions to the PVP and non-polarized sectors as λ is reduced. The requirement of only a small suppression λ~0.9 to reach the non-polarized phase ($\nu_+ = -\frac{2}{3}$ and $\nu_- = -\frac{2}{3}$) again points to the close energetic competition between distinct valley polarization sectors. Within the non-polarized sector, we do not observe signatures of a CDW at $\nu = -\frac{4}{3}$ in the numerical result. In the inset of Fig. 4c, we further assess the possibility of an FTI ground state and find that it is only present for θ ≤ 4.0°. Interestingly, this upper limit on the twist angle, which can be traced to the stability of the valley-polarized FCI at $\nu = -\frac{2}{3}$, is similar to the threshold of the experimentally-observed AFM-like behavior. While the FTI only survives to λ~0.5, we emphasize that a more definitive determination of FTI stability is highly sensitive to detailed treatment of band-mixing, screening and short-range interaction effects [16,57], which we defer to future studies.

In summary, we present magnetic signatures for a putative FTI in $tMoTe_2$ at $\nu = -\frac{4}{3}$. This interpretation is supported by ED calculations revealing the extremely small energy difference ($\Delta E < 1$ meV) between the putative FTI ($\nu_{\pm} = -\frac{2}{3}$ and $\nu_{\mp} = -\frac{2}{3}$) and PVP states ($\nu_{\pm} = -1$ and $\nu_{\mp} = -\frac{1}{3}$), explaining the experimental observation of small magnetic field in inducing first-order phase transitions. Recent transport measurements on $tMoTe_2$ confirmed the presence of a resistive state at $\nu = -\frac{4}{3}$ [13,14], but it is not known if the measured longitudinal resistance is determined primarily by the FTI resistance. During the revision of this manuscript, we note the posting by Chou and Das Sarma who theoretically showed that two-terminal transport of edge modes is insufficient to identify the $\nu = -\frac{4}{3}$ FTI state [58]. These results call for future transport experiments with advanced geometries and imaging measurements with edge resolution to further establish the $\nu = -\frac{4}{3}$ FTI state.

## METHODs

### 1. Device Fabrication and Characterization

Van der Waals flakes used to assemble the heterostructure devices, including graphite, h-BN, and monolayer $MoTe_2$, were mechanically exfoliated onto oxygen-plasma-cleaned $Si/SiO_2$ substrates and identified by optical contrast microscopy. Atomic force microscopy (AFM) was used to determine h-BN thickness and to verify that the flakes were free of surface residue. Owing to the air sensitivity of $MoTe_2$, its exfoliation and all subsequent fabrication steps were carried out in an argon-filled glovebox with $H_2O$ and $O_2$ levels below 0.1 ppm. Prior to assembly, the $MoTe_2$ monolayer was cut into two halves using an AFM tip. Heterostructures were fabricated using standard dry-transfer techniques. The top gate was assembled first by sequentially picking up h-BN, the top graphite electrode, the top-gate h-BN dielectric, and a graphite grounding pin. One half of the $MoTe_2$ flake was then picked up, while the second half was rotated by the desired twist angle before being picked up and stacked to form the moiré superlattice. The completed stack was subsequently placed onto a prefabricated back gate consisting of a graphite electrode capped with an h-BN dielectric, along with gold contacts and wire-bonding pads defined by electron-beam lithography and electron-beam evaporation to enable electrical access to both gates and the grounding pin. The back-gate surface was AFM-cleaned in contact mode prior to transfer. The final heterostructure was released onto the $SiO_2$/Si substrate ($SiO_2$ thickness: 285 nm for device 3.7°, 3.3° device and 90 nm for device 3.9° and 4.0°) by melting the PC stamp at approximately 170 °C, followed by dissolution of the polymer in anhydrous chloroform for 5 min inside the glovebox. For device D 3.7°, both the top- and bottom-gate h-BN dielectrics are 35 nm thick. For device D3.9°, the top-gate and bottom-gate h-BN thicknesses are 11.3 nm and 20.4 nm. For device D4.0°, the top-gate and bottom-gate h-BN thicknesses are 30 nm and 38 nm. For device D3.3°, the top-gate and bottom-gate h-BN thicknesses are 33 nm and 37 nm respectively. Graphite electrode thicknesses were estimated from optical contrast: the top-gate graphite corresponds to approximately 2–3 graphene layers and the bottom electrode is ~10 nm thick.

Each $tMoTe_2$ device was mounted in a closed-cycle cryostat (Quantum Design, OptiCool), with a base temperature reading of 1.65 K on the sample mount. While all temperatures values presented above were from the sample mount readings, the actual temperatures on the sample during experiments were likely higher.

## 2. Determination of Carrier Density and Twist Angle

The thickness of the bottom h-BN dielectric in twisted $MoTe_2$ ($tMoTe_2$) devices was measured using atomic force microscopy. The geometrical gate capacitance per unit area was calculated as $C = \varepsilon_{\text{h-BN}}\varepsilon_0/d_{\text{h-BN}}$, where $\varepsilon_{\text{h-BN}} = 3.0$ is the dielectric constant of h-BN, $\varepsilon_0$is the vacuum permittivity, and $d_{\text{h-BN}}$is the thickness of the corresponding h-BN layer. This expression was applied independently to the top and bottom gates to obtain the top-gate ($C_{\text{tg}}$) and bottom-gate ($C_{\text{bg}}$) capacitances. The carrier density in $tMoTe_2$ was determined according to $n = (V_{\text{tg}}C_{\text{tg}} + V_{\text{bg}}C_{\text{bg}})/e$, where $V_{\text{tg}}$and $V_{\text{bg}}$denote the applied top- and bottom-gate voltages, respectively, and $e$is the elementary charge. The twist angle of each device was extracted from photoluminescence spectroscopy. In particular, the exciton resonance exhibits pronounced and reproducible changes at superlattice filling factors $\nu = \pm 1$, which were used to determine the corresponding carrier densities. The twist angle was then calculated from the measured density using$\theta = \sqrt{8n/(\sqrt{3}a_0^2)}$, where $a_0 = 3.52$ Å is the lattice constant of $MoTe_2$.

### 3. Assignment of Filling Factors

The filling factor $\nu$ defines the number of holes (or electrons) per moiré superlattice site. To determine $\nu$, we use the well-established insulating states, $\nu = 0$, $\nu = \pm 1$ and $\nu = \pm 2$ to calibrate the respective conversion factor between gate voltage ($V_g = V_{tg} + V_{bg}$) and the filling factor. We identify the gate voltages for all insulating states, determined by the static PL spectrum for each device (Extended Data Fig. 1), further confirmed by the maximum change in reflection observed in the pump–probe maps (Fig 2).

For device D1 (θ = 3.9°), the gate voltages corresponding $\nu = 1$ and $\nu = -1$ are determined from the PL. Using a linear fit to these points, we establish the relationship between filling factor and gate voltage. The filling factors for other insulating states within this doping range are calculated from their gate voltages and the conversion factor. We focus on the hole-doped states. For hole-doped states beyond $\nu = -1$ the gate efficiency increases; therefore, we used $\nu = -2$ and $\nu = -1$ to define the conversion relationship for $-2 < \nu < -1$ states.

We apply a similar calibration procedure for the other devices, noting that D1 is the only device to reach filling factors beyond $\nu \geq -2$. We perform the assignment for D2 (θ = 3.7°), by linearly fitting the regions $-2/3 < \nu < 1$ and $-2/3 < \nu < -1$ from the PL separately to similarly account for increasing gate effectiveness at higher densities. The fractional states $\nu = -4/3$ ad $-3/2$ are extrapolated from the linear relationship of the $-2/3 < \nu < -1$ region. Due to limited resolution, the conversion factors for devices D3 (θ = 4.0°) and D4 (θ = 3.3°) are determined solely from linearly fitting the region $0 < \nu < -1$.

### 4. Pump–Probe Measurements

Ultrafast pump–probe measurements were performed using femtosecond laser pulses produced by a solid-state amplifier (CARBIDE, Light Conversion) operating at a repetition rate of 400 kHz, a central wavelength of 1,050 nm, and a pulse duration of 250 fs. The laser output was divided into pump and probe paths. For the probe arm, a fraction of the fundamental beam was focused into a YAG crystal to generate a stable white-light continuum, which was spectrally filtered using a 50 nm band-pass filter centered at ~1,110 nm to span the exciton and trion resonances of $MoTe_2$. In the pump arm, the beam was routed through a motorized delay stage to control the pump–probe delay $\Delta t$, then focused into a second YAG crystal to produce a broadband continuum that was subsequently filtered to 1,225 ± 50 nm. The pump beam was modulated by an optical chopper to alternate between pump-on and pump-off conditions. Pump and probe beams were combined collinearly and focused onto the sample using a 100× objective with numerical aperture 0.75, yielding spot diameters of approximately 1.5 μm for the pump and 1 μm for the probe. The effective pulse duration at the sample was estimated to be ~200 fs from the coherent artefact in the pump–probe cross-correlation.

The same objective was used to collect the reflected signal, which was spectrally filtered to suppress residual pump light and dispersed onto an InGaAs array detector (PyLoN-IR, Princeton Instruments). Transient reflectance spectra were obtained by comparing pump-on and pump-off signals at each delay time, and the differential reflectance was calculated as $\Delta R/R = [R(\Delta t) - R]/R$, where $R$denotes the reflectance without pump excitation. Note that, since pump-induced change is ≤1%, it makes little difference when $R$ in the denominator is replaced by that after pump.

A reflectance spectrum is characterized by derivative shape across a resonance, Fig. 1b. Transient reflectance monitors $\Delta R/R_0$, where $\Delta R = R(\Delta t) - R_0$; $R(\Delta t)$ and $R_0$ are the reflectance spectrum at the pump-probe delay of $\Delta t$ and without pump, respectively.

The pump photon energy was chosen below the optical gap of $MoTe_2$ to avoid direct excitation of excitons. Throughout the measurements, the pump fluence was varied from 7 to 42 μJ cm$^{-2}$, while the probe fluence was fixed at 22 μJ cm$^{-2}$ (Extended Data Fig. 10). The peak amplitude of the $\Delta R/R$signal increases linearly with pump fluence, and the characteristic electronic melting and recovery dynamics are independent of fluence over this range. Unless otherwise noted, all data presented in the main text were acquired using pump and probe fluences of 42 μJ cm$^{-2}$ and 22 μJ cm$^{-2}$, respectively**.**

**5. Time-Resolved MCD Measurements**

Time-resolved magnetic circular dichroism measurements were performed using the same pump–probe configuration described above. The pump beam conditions were kept identical, and its polarization was found not to affect the measured MCD response (Extended Data Fig. 2). A quarter-wave plate was inserted into the probe beam path to generate left- or right-handed circularly polarized probe light. For each measurement, the probe helicity was fixed while either the out-of-plane magnetic field or the sample carrier density was swept, yielding helicity-resolved maps as a function of magnetic field or doping. The MCD signal was obtained by taking the difference between the transient reflectance signals measured with left- and right-circularly polarized probe light. For magnetic-field-dependent measurements, the magnetic field was swept in both directions to account for and correct possible hysteresis effects.

**6. Exciton and Attractive Polaron Oscillator Strength.**

The optical spectral weight $f_{\mathrm{AP}}^{\pm}$ of the attractive polaron (AP) resonances in $\sigma^{\pm}$circular polarization is directly proportional to the hole density $n_{\mp}$ in the opposite $K_{\mp}$ valley. This proportionality enables AP resonances to serve as a direct optical probe of spin–valley polarization in twisted $MoTe_2$, as established in studies of exciton–polaron physics in two-dimensional semiconductors [59,60]. In these works, because resonant excitation of the AP transition does not involve intermediate energy relaxation processes, the circular polarization degree of the AP response,

$$\rho_{\mathrm{AP}} = \frac{f_{\mathrm{AP}}^{-} - f_{\mathrm{AP}}^{+}}{f_{\mathrm{AP}}^{-} + f_{\mathrm{AP}}^{+}} \tag{1}$$

directly reflects the spin–valley polarization of the hole system,

$$\langle S_z \rangle = \frac{n_{+} - n_{-}}{n_{+} + n_{-}} \tag{2}$$

We note that our experiment measures the transient differential reflectance $\Delta R/R$, which probes the *reduction* in the probe EX/AP oscillator strength $\Delta f_{\mathrm{AP}}^{\pm}$ pump-induced disordering at fixed time delays. In the limit where the correlated gap is fully disordered, the change in the probe oscillator strength will be directly proportional in magnitude to the equilibrium oscillator strength, i.e., $|\Delta f_{\mathrm{AP}}^{\pm}| \sim |f_{\mathrm{AP}}^{\pm}|$. We qualify this assumption with pump fluence dependence (Extended Data Fig 2), where the ΔR/R amplitude saturates at pump fluences below those used in the main text, consistent with complete disordering of the correlated state. Under this assumption, the transient optical response retains the same valley-selective proportionality to the carrier population as in equilibrium. Accordingly, we model the transient spectra using the same susceptibility formalism developed for static reflectance measurements, with all parameters understood to represent the instantaneous pump-modified states at a given time delay. For notational simplicity and clarity, we omit explicit Δ symbols on the fitted parameters.

Following the formalism developed in the previous works, the $\sigma^{\pm}$-polarized AP transitions in moiré $MoTe_2$ are described by a two-dimensional complex optical susceptibility,

$$\chi_{\mathrm{AP}}^{\pm}(E) = -\frac{\hbar c}{E_{\mathrm{AP}}^{\pm}} \frac{\gamma_{\mathrm{rad}}^{\pm}}{E - E_{\mathrm{AP}}^{\pm} + i\Gamma^{\pm}/2} \tag{3}$$

where $E$ is the photon energy, $E_{\mathrm{AP}}^{\pm}$ denotes the AP resonance energy, $\Gamma^{\pm}$ is the total decay rate, $\Gamma^{\pm} = \frac{\gamma_{\mathrm{rad}}^{\pm}}{\hbar} + \frac{\gamma_{\mathrm{nrad}}^{\pm}}{\hbar}$, and $\frac{\gamma_{\mathrm{rad}}^{\pm}}{\hbar}$and $\frac{\gamma_{\mathrm{nrad}}^{\pm}}{\hbar}$ are the radiative and non-radiative decay rates, respectively. The radiative decay rate is proportional to the oscillator strength $f_{\mathrm{AP}}^{\pm}$ and therefore encodes the valley-resolved carrier population.

The measured transient reflectance response is obtained by relating this two-dimensional susceptibility to the optical boundary conditions at the $MoTe_2$ sample. Specifically, the contribution of the AP transition to the complex reflection amplitude can be expressed as an effective sheet response by the Fresnel relation for a two-dimensional sheet,

$$r_{\mathrm{AP}}^{\pm}(E) = -\frac{i\omega}{2c}\, \chi_{\mathrm{AP}}^{\pm}(E) \tag{4}$$

with $\omega = E/\hbar$. To account for the multilayer environment of the sample, the total complex reflection amplitude is calculated using the transfer matrix method (TMM), which simulates the dielectric stack both with and without the $MoTe_2$ bilayer. In this framework, the entire moiré bilayer is treated as a two-dimensional conducting sheet whose optical susceptibility modifies the boundary conditions at its interface. The total reflection coefficient $R(E, \Delta t)$ in the presence of pump is obtained by inserting the sheet susceptibility into the transfer matrix of the full stack while the reference reflection coefficient $R_0(E)$is calculated for the same stack in the absence of the sample, defining the equilibrium reflectance measured at negative pump–probe delay (before time zero), with $R = \left|r_{tot}^{\pm}(E, \Delta t)\right|^2$. The differential reflectance is then evaluated as $\Delta R/R_0 = [R(E, \Delta t) - R_0(E)]/R_0(E)$.

For quantitative comparison with experiment, a smooth background contribution is included by adding a linear background term $s_{bg}(E - E_0)$ and constant offset $C$, accounting for residual reflectance from nearby optical transitions such as the neutral exciton or repulsive polaron, which are not explicitly included in $\chi_{\mathrm{AP}}^{\pm}$. In some cases, reflectance model is scaled by an amplitude term $A$ to account for possible losses in the beam path. This treatment follows the analysis approach adopted in references [59,60]. Extended Data Fig. 11 shows representative fitting at all four twist angles.

In the magnetic field dependent measurements at $\nu = -4/3$, the AP spectrum for a single probe polarization ($\sigma^+$ or $\sigma^-$) was iteratively fit using the model described above at each magnetic field step. Prior to fitting, regions of reduced contrast corresponding to the "zero" signal regions were subtracted from the 2D map. These regions exhibit no distinct spectral features and are essentially flat, leading to unphysical fit parameters. In any case, these spectral regions are subtracted away in the CD. The extracted oscillator strength $\Delta f_{\mathrm{AP}}^{\pm}$ as a function of magnetic field is shown in main text Figs. 2 and 3.

## 7. Optical Switching Effect

Optical switching of the Chern number for integer and fractional Chern Insulators has recently been demonstrated in moiré $MoTe_2$ [23,26,27]. We consider the possibility of resonant probe-induced switching as an alternate explanation for the AFM-like signature. We compare the net photon fluence in our transient experiments versus the experimental conditions in which switching has been reported. Holtzmann and coworkers achieve Chern number switching by using a supercontinuum laser (50 MHz) for several seconds on the sample at an average power of 100 nW, resulting a pulse energy of $10^{-9}$ μJ. Assuming their beam radius is on the order of ~μm, this corresponds to photon flux $\Phi_{\mathrm{switch}} \sim 10^{17}$ $\mathrm{cm^{-2}s^{-1}}$. A conservative estimate of 10 s integration to reach the switching threshold corresponds to a net photon fluence of $10^{18}$ photons/$\mathrm{cm}^2$. In our experiment, the probe pulse energy is $1.5\mathrm{x}10^{-7}$ μJ (from an average power *P*~200 nW and a laser repetition rate 400 kHz). Since the pump-probe experiment re-initiates the state for every pump-probe pulse pair, the net photon fluence, $10^{14}$ photons/$\mathrm{cm}^2$, on the sample is four-orders-magnitude lower than what is needed to induce switching. We conclude that resonant probe induced switching of the Chern number does not occur in our measurement.

## 8. Theoretical Calculations

### 8.1. Interacting continuum model

To theoretically study the moiré valence bands of $tMoTe_2$, we adopt the continuum model [61], which we briefly summarize here following the conventions of Refs. [43,62]. Owing to strong Ising spin-orbit coupling, the low-energy valence states of $tMoTe_2$ are spin-valley locked, such that valley $K$ ($K'$) is associated with spin ↑ (↓). Henceforth, we use the term valley to refer to this spin-valley degree of freedom. We introduce a flavor index $\tau = +,-$ to label valley $K^+$, $K^-$ respectively. The continuum model for $\tau = +$ reads

$$H_{kin}^{+} = \begin{pmatrix} h_b(\boldsymbol{r}) & t(\boldsymbol{r}) \\ t^*(\boldsymbol{r}) & h_t(\boldsymbol{r}) \end{pmatrix} \quad (5)$$

where the matrix structure refers to layer $l = b,t$. The non-interacting Hamiltonian for valley $K'$ can be obtained using time-reversal symmetry. The intralayer term is

$$h_l(\boldsymbol{r}) = \frac{\hbar^2\nabla^2}{2m^*} + V_l(\boldsymbol{r}) \quad (6)$$

where $m^* = 0.6m_e$, $(-1)^t = 1$ and $(-1)^b = -1$. The moiré potential $V_l(r)$ and interlayer hopping $t(r)$ are

$$V_l(\boldsymbol{r}) = Ve^{-(-)^l i\psi} \sum_{i=1,2,3} e^{i\boldsymbol{g}_i\cdot\boldsymbol{r}} + c.c. \quad (7)$$

$$t(\boldsymbol{r}) = w \sum_{i=1,2,3} e^{-i\boldsymbol{q}_i\cdot\boldsymbol{r}} \quad (8)$$

The moiré wavevectors are $\boldsymbol{q}_1 = \frac{8\pi \sin\frac{\theta}{2}}{3a_0}(0,1)$ and $\boldsymbol{q}_j = \hat{C}_3^{j-1}\boldsymbol{q}_1$, where $a_0 = 0.352$nm is the lattice constant of $MoTe_2$ and $\hat{C}_3$ denotes counterclockwise rotation by $2\pi/3$. We have also defined $\boldsymbol{b}_j = \boldsymbol{q}_{j+2} - \boldsymbol{q}_{j+1}$. Note that we keep only the first harmonics in the moiré terms. We use the parameters $w = -18.8$ meV, $V = 16.5$ meV, $\psi = -105.9^\circ$ which were extracted from *ab initio* calculations at $\theta = 3.89^\circ$. [62] When investigating different twist angles, we will simply change $\theta$ in the continuum model while keeping the other parameters fixed, which we expect to be a reasonable approximation in the range $\theta \simeq 3.4^\circ - 4.4^\circ$.

Diagonalizing the continuum model yields moiré Bloch wavefunctions and the single-particle dispersion indexed by the band label $m$ and the moiré momentum $\boldsymbol{k}$ in the moiré Brillouin zone (mBZ). We incorporate long-range density-density interactions as

$$\hat{H}_{int} = \frac{1}{2\Omega} \sum_{\boldsymbol{q}} \sum_{\tau\tau'} V_{\tau\tau'}(\boldsymbol{q}) : \hat{\rho}_\tau(\boldsymbol{q}) \hat{\rho}_{\tau'}(-\boldsymbol{q}) :_{\nu=0} \quad (9)$$

$$\hat{\rho}_\tau(\boldsymbol{q}) = \sum_{\boldsymbol{k}\in mBZ} \sum_{mn} \lambda_{mn}^{\tau}(\boldsymbol{k},\boldsymbol{q}) c_{\tau,\boldsymbol{k},m}^{\dagger} c_{\tau,\boldsymbol{k}+\boldsymbol{q},n} \quad (10)$$

$$\lambda_{mn}^{\tau}(\boldsymbol{k},\boldsymbol{q}) = \langle u_{\boldsymbol{k},m}^{\tau} | u_{\boldsymbol{k}+\boldsymbol{q},n}^{\tau} \rangle, \quad (11)$$

where $c_{\tau,\boldsymbol{k},m}^{\dagger}$ is an electron creation operator, $|u_{\boldsymbol{k},m}^{\tau}\rangle$ is the cell-periodic part of the moiré Bloch function, and $\Omega$ is the system area. $:\ldots:_{\nu=0}$ denotes normal-ordering with respect to the fully occupied valence bands — this is effectively equivalent to placing all electron creation operators to the right of annihilation operators. The interaction potential $V_{\tau\tau'}(q)$ used in the calculations will be specified below, though generically we have $V_{++} = V_{--}$ and $V_{+-} = V_{-+}$ due to time-reversal symmetry. The interacting continuum model consists of the non-interacting continuum model for both valleys augmented with the interaction $\widehat{H}_{int}$. Note that the Hamiltonian satisfies a valley-$U_v(1)$ symmetry, corresponding to conservation of particle number in each valley sector.

8.2. **Exact diagonalization calculations of spin polarization at $\nu = -4/3$**

*Setup*. To study the energy competition between different valley polarization sectors at $\nu = -4/3$ in $tMoTe_2$, we employ exact diagonalization (ED) calculations on the interacting continuum model. Owing to the $U_v(1)$ symmetry, this can be achieved by performing calculations at different partial valley fillings $\nu_+, \nu_-$, where $\nu_+ + \nu_- = \nu = -4/3$.

For the interaction potential, we use the dual gate-screened interaction

$$V_{\tau\tau'}(\boldsymbol{q}) = \frac{e^2}{2\epsilon_0\epsilon_r q}\tanh\frac{q\xi}{2}, \tag{12}$$

with dielectric constant $\epsilon_r = 10$ and gate-screening parameter $\xi = 20$nm. Since $V_{\tau\tau'}(q)$ represents the 'bare' longrange Coulomb interaction, it is valley isotropic in that it does not depend on the valley indices $\tau, \tau'$. This is motivated from the fact that the two valleys in $MoTe_2$ occupy the same region in real-space.

Due to the exponential complexity of ED, it is necessary to reduce the single-particle Hilbert space, which is achieved in two ways. First, we consider small finite systems with $N_s$ moiré unit cells. Second, we perform band truncation and restrict our attention to the lowest two valence bands per valley, which are referred to as band0 and band1 (see Extended Data Fig. 12a). This implies that all other valence bands are assumed to be fully occupied, i.e. devoid of holes. However, even for a small system with $N_s = 12$, the momentum-resolved many-body Hilbert space dimension for a two-band calculation at $\nu_+ = \nu_- = -2/3$ is $\sim 3 \times 10^{10}$, which is prohibitively large. Therefore, we further implement the 'band-max' truncation technique previously utilized in Refs. [16], [63]. In the current context, this involves starting with the limit of a single-band calculation ($N^1_{max} = 0$) where the holes are only allowed to enter band0, while band1 is fully occupied by electrons. However, it has been established from prior theoretical studies that the effects of band1 are important for capturing even qualitative aspects of the phase diagram [55,56,62]. This is not surprising given that the indirect gap between band0 and band1 is small. To systematically approach the two-band limit, we thus perform a sequence of ED calculations where we allow at most $N^1_{\max}$ holes to populate band1. Intuitively, this enables 'band-mixing' between band0 with the next valence band.

In the single-band limit, the maximum allowed polarization is the partially valley-polarized (PVP) sector where $\nu_+ = -1$ and $\nu_- = -1/3$. Therefore, in our band-mixing calculations, we focus primarily on the competition between the PVP sector and the non-polarized (NP) sector with $\nu_+ = \nu_- = -2/3$. We compute the ground state energies $E_{\mathrm{PVP}}$ and $E_{\mathrm{NP}}$ in these sectors and compare their difference as a function of twist angle and band-mixing $N^1_{max}$.

*Results.* In Figs. 1b)-f), we plot $E_{\mathrm{PVP}} - E_{\mathrm{NP}}$ per unit cell as a function of $\theta \in [3.4^\circ, 4.2^\circ]$ and $N^1_{max}$ for different finite lattices. Note that in order to gauge the finite-size effects, for $N_s = 12, 15$, we have used tilted lattices to generate different momentum meshes (inset) for the same $N_s$. These tilted lattices have more isotropic aspect ratios than their non-tilted counterparts. For all meshes considered, the PVP sector has lower energy than the NP sector in the single-band limit ($N^1_{max} = 0$). However, the energy difference is small, being less than 1meV per unit cell, implying a close energetic competition. As we increase $N^1_{max}$, the energy of the NP sector is lowered relative to the PVP sector. For the $N_s = 12$ meshes [see panels b) and c)], we observe that the NP sector attains a lower energy than the PVP sector, though it is possible that this also eventually occurs for the $N_s = 15$ meshes if $N^1_{max}$ is pushed beyond the limit we can currently access numerically. Our results strongly suggest that band-mixing has the effect of depolarizing the system (see also Ref. [2]), though the energetic competition remains very close.

Due to finite-size effects, we do not characterize the nature of the ground state within each valley polarization sector in these small-$N_s$ calculations. For example, even the stability of the $\nu = -2/3$ valley-polarized fractional Chern insulator (FCI) is sensitive to finite-size effects for such small $N_s$. We observe noticeable finite-size variations in the results of Extended Data Fig. 12. We are hence unable to extract a definite twist angle dependence of the spin polarization at $\nu = -4/3$ from Extended Data Fig. 12.

Finally, we consider the fully valley-polarized (FVP) sector ($\nu_+ = -4/3$ and $\nu_- = 0$). Note that the FVP sector necessarily involves a non-zero number of holes beyond band0, and hence cannot be accessed without band-mixing. For this reason, we are forced to restrict to the small system $N_s = 9$ where the calculation is computationally feasible across the FVP, PVP, NP sectors for sufficiently large $N_{max}^{1}$. Our results in Extended Data Fig. 12g) suggest that the FVP sector is less competitive than the PVP sector. We remark that a previous experimental investigation found that large magnetic fields $\sim$ 10T are required to fully valley-polarize the system at $\nu = -2$. [25]

### 8.3. **Investigation of $n_s = 36$ system at $\nu = -4/3$**

*8.3.1. Setup.*

Motivated by the finite-size effects of the unbiased ED calculations in Sec. II, we investigate a larger system size of $N_s = 6 \times 6$ in this section. Such a lattice avoids geometrical frustrations that can artificially penalize CDW states at the $K_M$, $K_M'$ or $M_M$ points. While a larger $N_s$ is expected to reduce the finite-size effects, it also necessitates more constraining approximations to tame the growth of the Hilbert space. For example, even a single-band ED calculation with $N_s = 36$ for the NP sector involves a many-body Hilbert space of dimension $\sim 4 \times 10^{16}$. In the following, we describe our approach to obtain 'model-state variational' energies in the FVP, PVP, and NP sectors for twist angles $\theta \in [3.4^\circ, 4.4^\circ]$.

*8.3.2. Model-state variational calculation for the NP sector.*

For the NP sector ($\nu_+ = \nu_- = -2/3$), the computational difficulty arises because holes occupy both valleys. However, a fully valley-polarized single-band calculation at $\nu = -2/3$ only involves a Hilbert space dimension of $3.5 \times 10^7$, which is feasible. Furthermore, several of the candidate states at $\nu = -4/3$ can be adiabatically connected to a decoupled product of many-body states from the two valleys. For example, a model wavefunction for the fractional topological insulator (FTI) at $\nu = -4/3$ can be constructed via the decoupled product of $\nu_\tau = -2/3$ FCIs from the two valleys.

The above discussion motivates the following 'model-state variational' calculation to study the NP sector at $\nu = -4/3$. First, we perform a valley-polarized single-band ED calculation at $\nu_+ = -2/3, \nu_- = 0$. From this, we select some number of lowest energy many-body states $|\phi_\alpha^+\rangle$ indexed by $\alpha$, with corresponding many-body energies $E_\alpha^+$. The choice of states will be discussed shortly. By time-reversal symmetry, we can obtain analogous results for $\nu_+ = 0, \nu_- = -2/3$, yielding $|\phi_{\bar{\beta}}^-\rangle$ and $E_\beta^-$. We will refer to these selected states as 'input states'. To address the NP sector with $\nu_+ = \nu_- = -2/3$, the idea then is to consider a variational calculation in the space of many-body states $|(\alpha,\beta)\rangle \equiv |\phi_\alpha^+\rangle \otimes |\phi_{\bar{\beta}}^-\rangle$, spanned by taking products of the input states. Note that the total many-body Hamiltonian can be decomposed as

$$\hat{H} = \hat{H}^+ + \hat{H}^- + \hat{H}^{inter} \tag{13}$$

$$\hat{H}^{\tau} = \hat{H}^{\tau}_{kin} + \hat{H}^{\tau}_{int} \tag{14}$$

$$\hat{H}^{\tau}_{int} = \frac{1}{2\Omega}\sum_{\boldsymbol{q}} V_{\tau\tau}(\boldsymbol{q}) \colon \hat{\rho}_{\tau}(\boldsymbol{q})\hat{\rho}_{\tau'}(-\boldsymbol{q}) \colon_{\nu=0} \tag{15}$$

$$\hat{H}^{inter} = \frac{1}{\Omega}\sum_{\boldsymbol{q}} V_{+-}(\boldsymbol{q}) \colon \hat{\rho}_{+}(\boldsymbol{q})\hat{\rho}_{-}(-\boldsymbol{q}) \colon_{\nu=0}, \tag{16}$$

where we have used the fact that $V_{+-} = V_{-+}$. The goal is to construct the effective Hamiltonian with matrix elements

$$\tilde{H}_{(\alpha,\beta);(\alpha',\beta')} = \langle(\alpha,\beta)|\hat{H}|(\alpha',\beta'\rangle. \tag{17}$$

Clearly, we have

$$\langle(\alpha,\beta)\big|\hat{H}^{+} + \hat{H}^{-}\big|(\alpha',\beta'\rangle = \delta_{\alpha\alpha'}\delta_{\beta\beta'}(E^{+}_{\alpha} + E^{-}_{\beta}) \tag{18}$$

The remaining intervalley contribution $\langle(\alpha,\beta)|\hat{H}^{inter}|(\alpha',\beta'\rangle$can be assembled via the valley-diagonal one-body correlation functions $\langle\phi^{+}_{\alpha}|\hat{\rho}_{+}(\boldsymbol{q})|\phi^{+}_{\alpha'}\rangle$ and $\langle\phi^{-}_{\beta}|\hat{\rho}_{-}(\boldsymbol{q})|\phi^{-}_{\beta'}\rangle$ of the input states. We can then diagonalize $\tilde{H}$.

We now discuss the choice of input states arising from the $\nu_{+} = -2/3, \nu_{-} = 0$ calculation. Two possibilities for incompressible phases at $\nu_{+} = -2/3, \nu_{-} = 0$ include an FCI and charge density waves (CDW). On the 6 × 6 lattice, these manifest as distinct signatures in the low-energy many-body spectrum. In particular, an FCI would yield a three-fold topological degeneracy at total momentum $(k_1,k_2) = (0,0)$ separated by a finite gap to higher states. On the other hand, a $K_M$-CDW corresponding to a $\sqrt{3} \times \sqrt{3}$ unit cell reconstruction would show up as three degenerate ground states at distinct momenta $(k_1,k_2) = (0,0),(2,2),(4,4)$ separated by a finite gap to higher states. We choose 74 input states corresponding to the four (two) lowest states for zero (non-zero) total many-body momentum, which allows both these possibilities to appear. This means that the corresponding effective Hamiltonian $H^{\sim}$ for the NP sector at $\nu = -4/3$ has an effective Hilbert space dimension of approximately $74^2/36 \simeq 150$ per momentum, which can be straightforwardly diagonalized. We have checked that retaining a smaller number of input states [e.g. just the 3 lowest states at $(k_1,k_2) = (0,0)$] does not qualitatively affect the lowest energy of $\tilde{H}$.

In Extended Data Fig. 13a), we plot the many-body energies of the input states for the $\nu_{+} = -2/3, \nu_{-} = 0$ calculation for some representative angles, and the interaction potential of Eq. 12 (note that the intervalley part $V_{+-}$ does not matter for $\nu_{+} = -2/3, \nu_{-} = 0$). For $3.4^{\circ} \leq \theta \leq 4.0^{\circ}$, we find clear evidence for a $\nu = -2/3$ FCI, as demonstrated by the three-fold topological degeneracy at $(k_1, k_2) = (0,0)$. A possible candidate for an incompressible state in the NP sector at $\nu = -4/3$ is hence an FTI. For angles greater than 4.0°, the FCI becomes destabilized. However, for the whole twist angle range studied, we do not find any evidence for a $K_M$-CDW or any other charge order. This is suggestive that a charge-ordered phase is not a competitive candidate in the NP sector at $\nu = -4/3$.

To further rule out the scenario of a $K_M$-CDW candidate in the NP sector at $\nu = -4/3$, we perform additional $\nu = -5/3$ ED calculations at $\nu_+ = -2/3, \nu_- = -1$, where the $\tau = -$ sector simply corresponds to a fully hole-occupied band0. The rationale is that $\nu = -4/3$ is closer to $\nu = -5/3$ than $\nu = -2/3$, and the $\nu = -5/3$ calculation allows us to test whether the interaction-induced potential generated by the filled band0 in $\tau = -$ can drive CDW formation in $\tau = +$. Note that in these $\nu = -5/3$ calculations, we use the valley-isotropic interaction of Eq. 12. We find that the FCI survives for $3.4^\circ \leq \theta \leq 4.0^\circ$, and there are no signatures of $K_M$-CDW formation.

We defer a discussion of the intervalley interaction $V_{+-}(\boldsymbol{q})$ to Sec. 7.3.6, and the results of the diagonalization of $H^{\sim}$ to Sec. 8.3.7.

*8.3.3. Model-state variational calculation for the PVP sector*

For the PVP sector ($\nu_+ = -1, \nu_- = -1/3$), we perform an analogous variational calculation as for the NP sector in Sec. 8.3.2. The difference is that there is only a single input state for valley $K$, corresponding to fully populating band0 with holes. For valley $K'$, we generate the input states by performing a valley-polarized single-band ED calculation at $\nu = -1/3$. Two possibilities for incompressible phases at this filling include an FCI and a $K_M$-CDW, which have the same signatures in the many-body spectrum as the corresponding phases at $\nu = -2/3$. Therefore, we choose 74 input states corresponding to the four (two) lowest states for zero (non-zero) total many-body momentum.

In Extended Data Fig. 13b), we plot the many-body energies of the input states for the $\nu = -1/3$ calculation for some representative angles, and the interaction potential of Eq. 12 (note that the intervalley part $V_{+-}$ does not matter for $\nu_+ = -1/3, \nu_- = 0$). For smaller angles $\theta \leq 3.5^\circ$, we find evidence of an FCI. However, for $\theta \geq 3.7^\circ$, we instead find signatures of a $K_M$-CDW. This suggests that both FCI and $K_M$-CDW are candidate phases for the PVP sector at $\nu = -4/3$.

*8.3.4. Hartree-Fock calculations allowing for IVC*

The ED calculations naturally capture phases with definite valley polarization, but are not as well suited to probe states with $U_v(1)$ symmetry-breaking, such as intervalley-coherent (IVC) phases. To complement the ED calculations in the NP and PVP sectors, we therefore perform Hartree-Fock (HF) calculations on the same $N_s = 6\times6$ mesh within band0. In the HF computation, we allow for IVC at intervalley wavevectors $q = \Gamma_M, K_M, K_M'$ .

*8.3.5. Restricted calculation for the FVP sector*

While the FVP sector ($\nu_+ = -4/3, \nu_- = 0$) does not suffer from the complexity of having holes in both valleys, it does necessarily require the presence of holes beyond band0. A full two-band ED calculation in the FVP sector with $N_s = 36$ involves a Hilbert space of $\sim 2 \times 10^{17}$. To obtain a manageable calculation, we perform a restricted calculation where band0 is enforced to be fully populated with holes, and the remaining holes enter band1. This leads to a reduced Hilbert space of $\sim 3.5 \times 10^7$.

*8.3.6. Intervalley interaction*

We now discuss the form of the interaction used for the calculations on the $N_s = 6 \times 6$ lattice. For the intravalley interaction, we use the same dual-gate-screened form $V_{++}(q) = V_{--}(q) = \frac{e^2}{2\epsilon_0\epsilon_r q}\tanh\frac{q\xi}{2}$ with $\epsilon_r = 10$ and $\xi = 20$nm, as in Sec. 8.2. The physics of the FVP sector is not affected by the intervalley interaction. For the other calculations in this section though, the restriction to band0 means that band-mixing effects have been neglected. In Sec. 8.2., we found that a key consequence of band-mixing is the tendency to depolarize the ground state (see also Ref. [62]). To incorporate this in our band0-projected calculations in a simple manner, we introduce a parameter $\lambda$ that artificially tunes the strength of the intervalley interaction according to

$$V_{+-}(q) = \lambda V_{++}(q). \tag{19}$$

As demonstrated later in Sec. 8.3.7, this will relatively lower the energy of sectors with small valley polarization. In fact, the $q = 0$ component of the intervalley interaction alone leads to an energy shift of $-\frac{V_{+-}(0)\delta N^2}{\Omega}$, where $\delta N$ is the difference in number of holes between valley $K$ and $K'$.

Suppression of the positive quantity $V_{+-}(0)$ will therefore relatively favor the NP sector with $\delta N = 0$.

Refs. [57], [64] have previously employed second-order perturbation theory to study inter-band effects in a model consisting of Landau levels with opposite magnetic fields in the two valleys. They found that interband processes tend to weaken the intervalley interaction compared to the intravalley interaction. Ref. [16] also introduced a phenomenological short-range attraction that is only operative in the intervalley channel. In the following, we will simply treat $\lambda$ as a phenomenological tuning parameter and leave a more microscopic treatment to future work.

*8.3.7. Results*

In Extended Data Fig. 14a), we show the energy competition between the PVP and NP sectors as a function of $\lambda$ and $\theta$. We find that a suppression of the intervalley interaction ($\lambda < 1$) can lead to a transition from the PVP to the NP sector. The required intervalley suppression is quite small as the transition occurs for $\lambda \sim 0.9$ for the twist angle range studied, pointing to the close competition between the different valley polarization sectors. We note that the phase boundary moves to higher $\lambda$ for larger twist angles. However, we caution that $\lambda$ is a phenomenological parameter so that we refrain from making definite conclusions regarding the twist angle dependence, especially given the approximations involved in the calculations.

Extended Data Fig. 14b) shows analogous behavior for the energy competition between the FVP and NP sectors. Extended Data Fig. 14c) summarizes the ED phase diagram of the ground state polarization. Starting from the FVP sector in the valley-isotropic limit $\lambda = 1$ of the interaction, the ground state depolarizes as a function of $\lambda$ until it reaches the NP sector.

In Extended Data Fig. 14d), we examine the spectrum of projected Hamiltonian $\widetilde{H}$ in the NP sector to diagnose the presence of an FTI. An FTI manifests as a 9-fold topological degenerate ground state at $(k_1,k_2) = (0,0)$, separated by a finite gap to higher states. We define the FTI spread as the energy difference between the ninth-lowest and lowest energy states at $(k_1,k_2) = (0,0)$. We also define the FTI gap as the energy difference between the lowest state not in this set of nine states, and the ninth-lowest energy state at $(k_1,k_2) = (0,0)$. The FTI spread/gap is then the ratio of these two quantities and should be small for a well-formed FTI. The FTI is only present for twist angles less than $4.0^\circ$. For $\lambda = 0$, this can be understood from the fact that the two valleys are decoupled, so that the FTI is only present if the $\nu = -2/3$ calculation yields an FCI. We find that the FTI remains stable up to $\lambda \sim 0.5$. We note that the stability of the FTI is known to be highly sensitive to modelling details of both the short-range interaction (not included here) and the long-range interaction [5]. We do not find any signatures of a $K_M$-CDW in the NP sector.

In Extended Data Figs. 14e,f), we plot the average valley polarization $\langle S_z \rangle$ and the magnitude of IVC in the HF calculations. Recall that because the HF calculations are restricted to band0, that the maximal valley polarization is $\langle S_z \rangle = 2/3$. Similar to the ED calculations in Extended Data Fig. 13a), we find a transition between a partially polarized phase and a phase with no net polarization for a weak intervalley suppression $\lambda \sim 0.9$. We also observe a region of IVC for moderate $\lambda$ within the region of no average valley polarization. For smaller $\lambda \lesssim 0.5$, the IVC disappears (not shown).

### 8.4. Exact diagonalization calculations of spin polarization at $\nu = -3/2$

In this section, we follow the same approach as in Sec. 8.2, except applied to study the valley polarization at $\nu = -3/2$. Extended Data Fig. 15 shows the energy competition between the PVP ($\nu_+ = -1$ and $\nu_- = -1/2$) and NP ($\nu_+ = \nu_- = -3/4$) sectors on the $N_s = 16$ lattice. Analogous to the $\nu = -4/3$ case, we find that the PVP sector has lower energy for $N^1_{max} = 0$. The energy difference per unit cell is smaller than at $\nu = -4/3$ (see Extended Data Fig. 12). Incorporating band-mixing relatively favors the NP sector, and we anticipate that larger $N^1_{max}$ may lead to the NP sector being lower energy across all twist angles. Note that the rapid growth of the Hilbert space dimension ($2.1\times10^5$, $1.6\times10^7$, $5.4\times10^8$ for $N^1_{max} = 0,1,2$ respectively) prevents us from studying larger $N^1_{max}$. The small system size also prevents us from reliably analyzing the nature of the many-body ground state.

We note that HF calculations in the NP sector on larger system sizes can find a gapped $2 \times 2$ translation symmetry-breaking $N_s$ = 4x4 state, but further theoretical work is needed to fully understand the phase diagram at $\nu = -3/2$.

**ACKNOWLEDGEMENTS**

XYZ acknowledges support for GEM by the Department of Defense Multidisciplinary University Research Initiative grant number W911NF2410292 and for YW by the US Department of Energy, Office of Basic Energy Sciences (DOE-BES) under award DE-SC0024343, both for the pump-probe experiments. XX acknowledges support by the U.S. Department of Energy (DOE), Office of Science, Basic Energy Sciences (BES), under the award DE-SC0018171 for the fabrication of devices D1, D3, and D4 and the Vannevar Bush Faculty Fellowship, Award number N000142512047, for the fabrication of device D2. JHC acknowledges support for the bulk $MoTe_2$ crystal growth and characterization and JCH acknowledges support for sample characterization, by Programmable Quantum Materials, an Energy Frontier Research Center funded by DOE BES under award DE-SC0019443. JI is supported by the National Science Foundation grant DMR-2011738 (to XYZ and RQ) through the Columbia University Materials Research Science and Engineering Center (MRSEC) on Precision-Assembled Quantum Materials. The Flatiron Institute is a division of the Simons Foundation. Additional support is acknowledged by the US Army Research Office grant number W911NF-23-1-0056 (XYZ and XR) for partial support of the experimental setup. K.W. and T.T. acknowledge support from the JSPS KAKENHI (Grant Numbers 21H05233 and 23H02052) and World Premier International Research Center Initiative (WPI), MEXT, Japan.

**Author Contributions**

X.Y.Z., X.X., Y.W., and G.E.M. conceived this work. Y.W. and G.E.M. conducted all spectroscopic measurements, analyzed, and interpreted the results, with the assistance of J.C. under the supervision of X.Y.Z. and X.R. Sample fabrication was carried out by W.L., S.Y., E. A., C. H. under the supervision of X.X. Y.K. and N.R. carried out all theoretical calculations and simulations. J.H.C. was responsible for $MoTe_2$ crystal growth. T.T. and K.W. provided the hBN crystal. J.I. contributed to the interpretation of optical selection rules, under the supervision of X.Y.Z. and R.Q. J. C.H. advised on sample preparation and characterization. The manuscript was prepared by Y.W., G. E.M., X.Y.Z., X.X., N.R. and Y.K., incorporating inputs from all coauthors. X.Y.Z. supervised the project. All authors read and commented on the manuscript.

**Competing Interests.** The authors declare no competing interests.

**Data Availability Statement. T**he data shown in the main figures are openly available [65]. Data supporting the findings of this study are available from the corresponding author upon request.

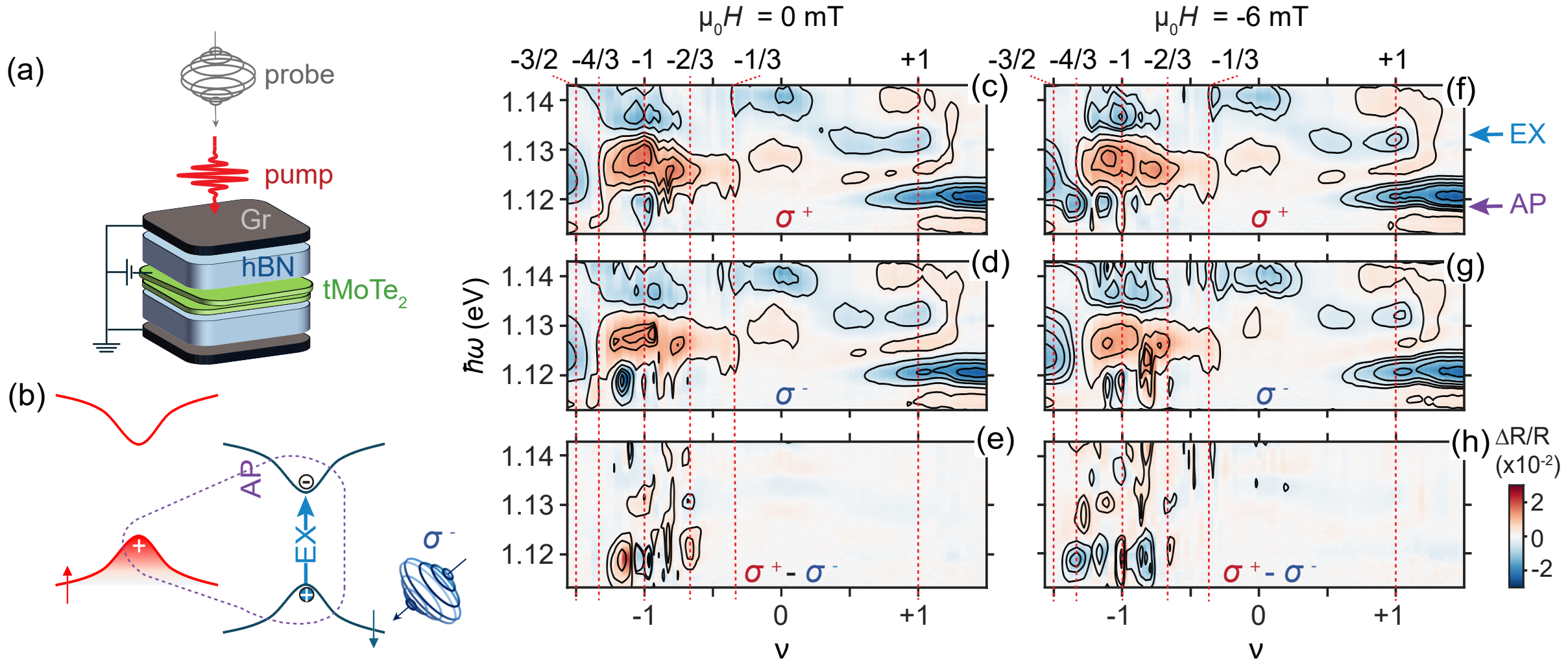

**Figure 1. Transient circular dichroism (CD) reveals distinct spectral signatures for the $\boldsymbol{\nu = -\frac{4}{3}}$ state.** (a) Illustration of the tCD experiment on dual-gated $tMoTe_2$ devices. A pump pulse with photon energy below the semiconductor band gap selectively disrupts correlated states in $tMoTe_2$. A probe pulse with circular polarization ($\sigma^+$ or $\sigma^-$) determines reflectance (R) at the exciton and trion resonances; (b) Schematic illustration of an exciton (EX) in the $K^-$ valley bound to a carrier in the polarized $K^+$ valley to constitute an attractive polaron (AP) transition excited by $\sigma^-$ light. (c-h) Spectrally resolved transient reflectance maps as a function of moiré filling factor (ν) for a $tMoTe_2$ device D1 (θ = 3.9°) at external magnetic fields of $\mu_0H = 0$ (c-e) and $\mu_0H = -6$ mT (f-h). The top and middle panels are ΔR/R spectra obtained with probe polarizations of $\sigma^+$ and $\sigma^-$, respectively, while the bottom panels present the difference spectra between the two probe polarizations ($\sigma^+$-$\sigma^-$) as proxy to CD spectra. The red-dashed lines mark filling factors of $\nu$ = -3/2, -4/3, -1, -2/3, -1/3, and +1. All spectra obtained at sample stage temperature of T = 1.6 K and a pump-probe delay of Δt = 1.2 ns.

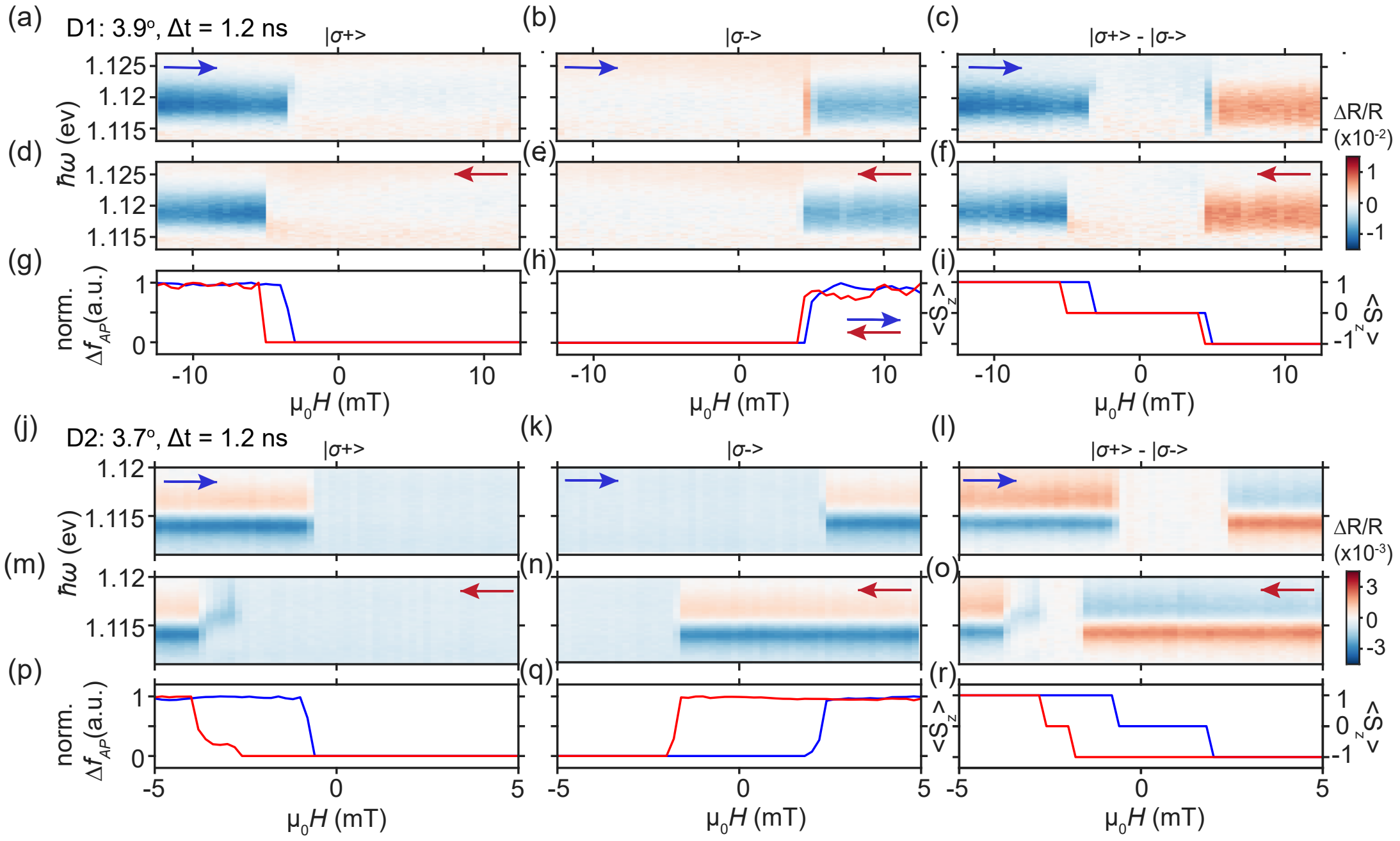

**Figure 2. AFM-like magnetic response of the $\boldsymbol{\upsilon = -\frac{4}{3}}$ state.** a-i Device D1 (θ = 3.9°). Transient reflectance spectra (colored scale ΔR/R) under forward magnetic field sweep probed by a) $\sigma^+$, b) $\sigma^-$, and c) $\sigma^+ - \sigma^-$ and backward magnetic field sweep probed by d) $\sigma^+$, e) $\sigma^-$, and f) $\sigma^+ - \sigma^-$. AP oscillator strength $\Delta f_{AP}^{\pm}$ extracted from the ΔR/R spectra under forward (blue) and backward (red) magnetic field sweeps probed by g) $\sigma^+$, h) $\sigma^-$ polarizations, and i) the corresponding valley/spin polarizations $< S_z > = (\Delta f_{AP}^{+} - \Delta f_{AP}^{-})/(\Delta f_{AP}^{+} + \Delta f_{AP}^{-})$. j-r) Similar data to a-i) for device D2 (θ = 3.7°). Attractive polaron oscillator strengths are extracted from modeling the 2D optical susceptibility described in Methods. All data obtained at Δt = 1.2 ns and nominal sample stage temperature T = 1.6 K.

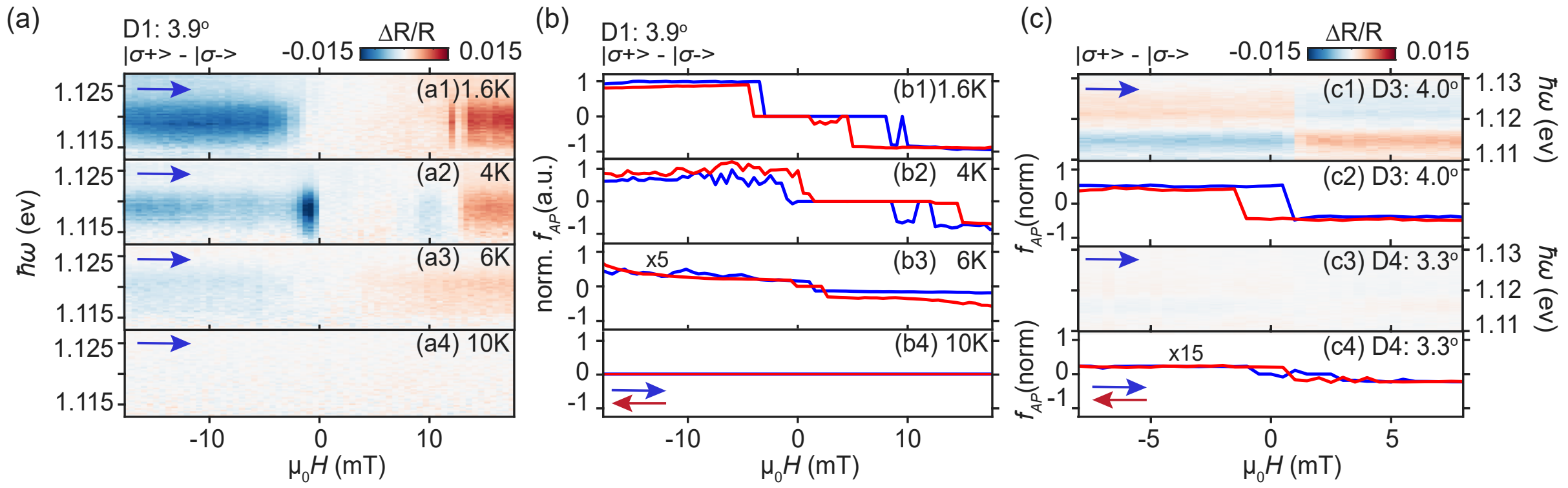

**Figure 3. Temperature and twist angle dependences in the magnetic behavior of the ν = -4/3 state.** a) CD color maps ($\sigma^{+}$-$\sigma^{-}$) for the forward sweeping magnetic field at T = 1.6, 4, 6, and 10K (a1-4). Signal has notably disappeared at 10K. b) corresponding CD in AP oscillator strength, ($\sigma^{+}$-$\sigma^{-}$), at the four temperatures: T = 1.6, 4, 6, and 10K (b1-4). Note that CD in AP oscillator strength in panel (b3) has been multiplied by a factor of 5; c) CD color maps ($\sigma^{+}$-$\sigma^{-}$) for the forward sweeping magnetic field and the corresponding AP oscillator strength for devices D3 (θ = 4.0°) and D4 (θ = 3.3°) at T = 1.6 K. The hysteresis profiles become ferromagnetic at θ = 4.0° and paramagnetic at θ = 3.3°. Note the CD in AP oscillator strength has been multiplied by a factor of 15. All transient spectra obtained at Δt = 1.2 ns.

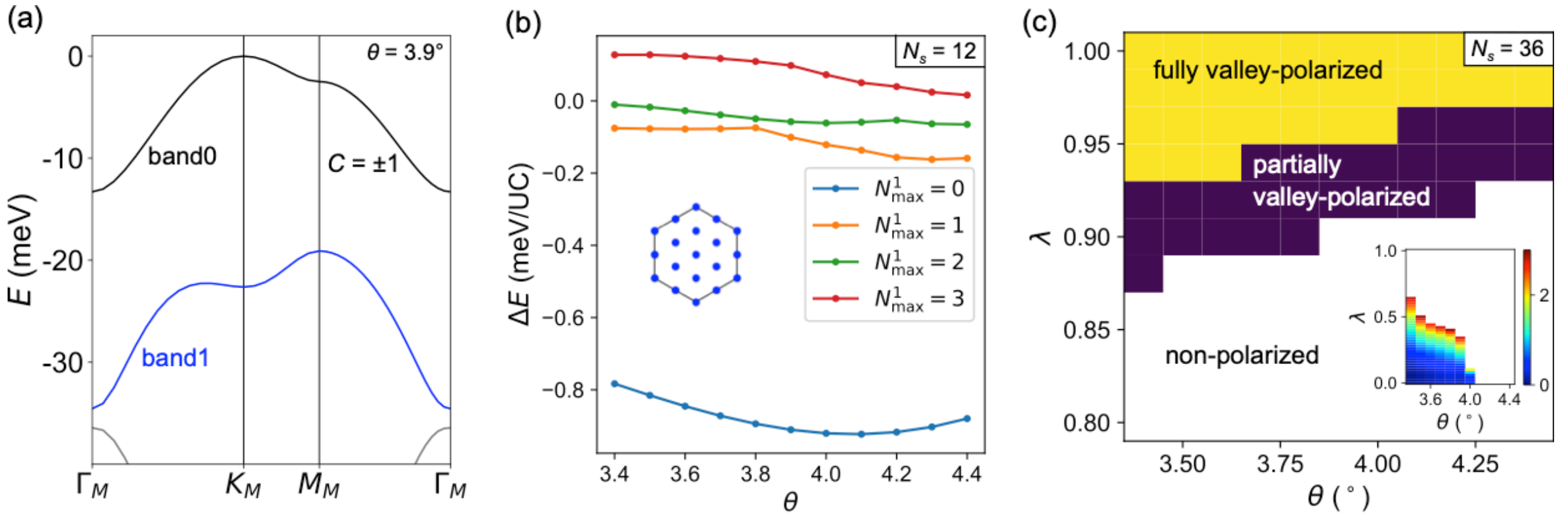

**Figure 4 Exact diagonalization calculations at ν = -4/3.** a) Continuum model band structure of the valence bands of $tMoTe_2$ at θ = 3.9°. The band closest to neutrality (band0) has valley Chern number |*C*|=1. b) Energy difference $\Delta E = E_{partial} - E_{non.}$ per unit cell between the partially valley-polarized (ν=-1, -1/3 ) and non-polarized (ν = -2/3, -2/3) sectors as a function of twist angle for the 12-site lattice (inset shows momentum mesh). The different curves correspond to the maximum number $N^1_{max}$ of holes allowed in the second valence band (band1). c) Ground state polarization in the variational ED calculation as a function of twist angle and intervalley interaction parameter λ for the 36-site lattice. Inset shows the FTI ratio, defined as the energy spread of the topological ground states divided by the gap. Non-white shading indicates a well-formed FTI.

# Extended Data

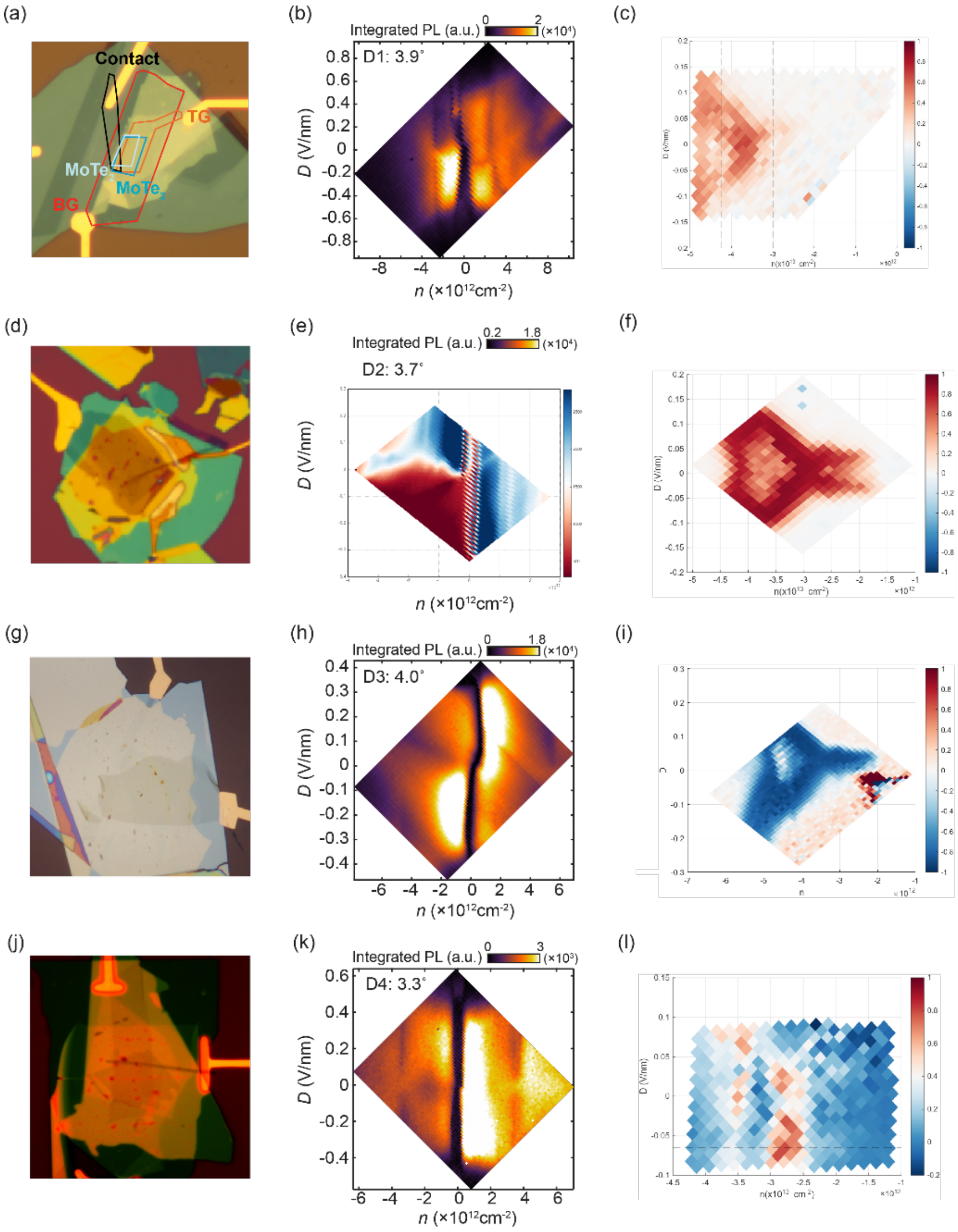

**Extended Data Fig 1**. Device images and characterization. a-c) D1 ($\theta$ = 3.9°), d-f) D2 ($\theta$ = 3.7°), g-i) D3 ($\theta$ = 4.0°), and j-l) D4 ($\theta$ = 3.3°). The first column presents optical images of the devices. The second column presents the trion PL for each device as a function of carrier density (n) and displacement field (D). The third column shows the n-and D-resolved degree of circular polarization, $\rho = \frac{PL(\sigma^+)-PL(\sigma^-)}{PL(\sigma^+)+PL(\sigma^-)}$. Integer and Red ($\rho$~1) corresponds to the ferromagnetic region. Fractional Chern insulating states at $\nu$ =-1 and $\nu$ =-2/3 are indicated by a reduction in $\rho$ contrast, and are used to calibrate the filling factor in the transient reflectance maps (Fig. 2, Extended Data Figs. 3, 7).

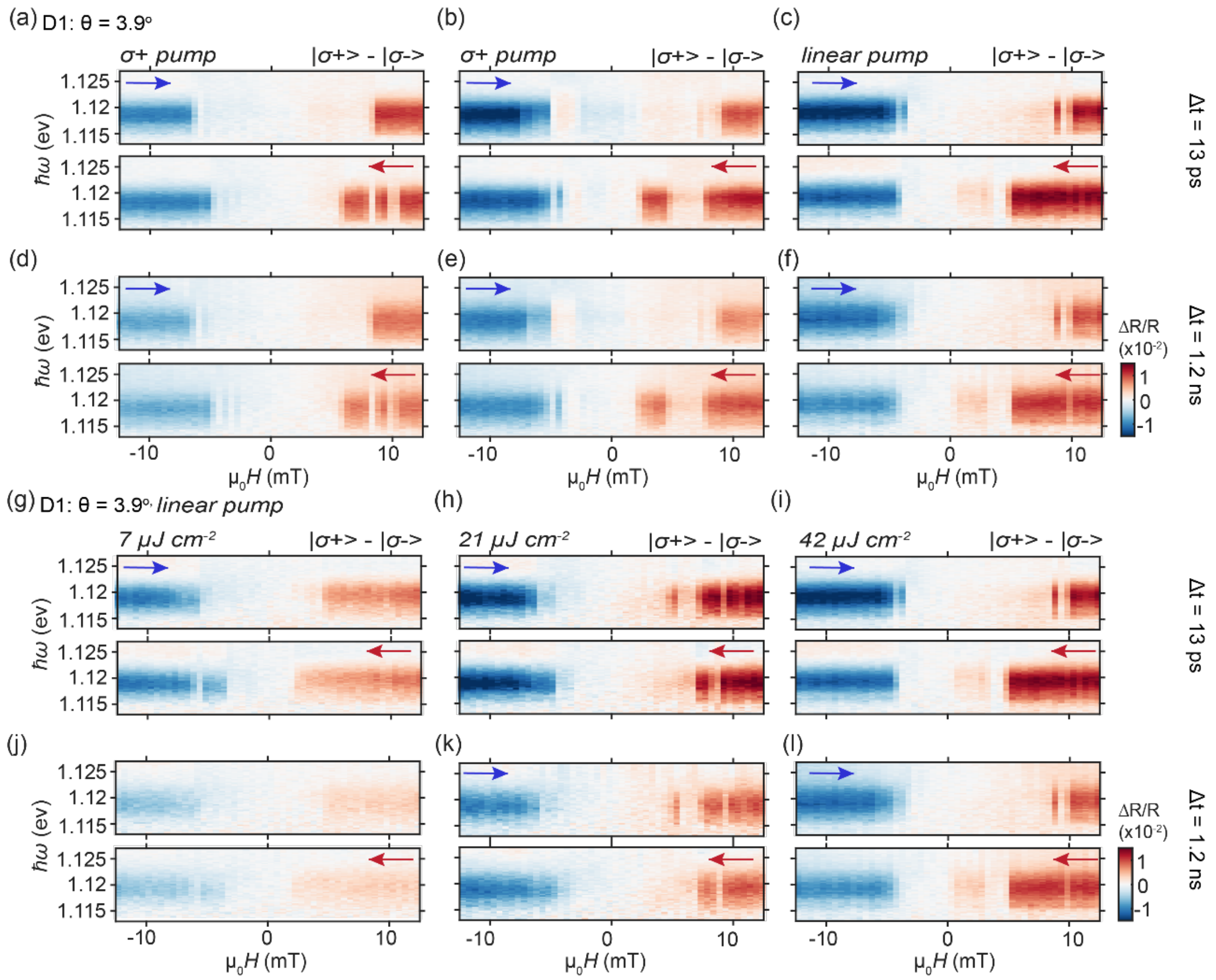

**Extended Data Fig 2. Pump polarization and fluence controls.** a-f Spectrally-resolved CD color maps ($\sigma^+$-$\sigma^-$) for the forward (top) and reverse (bottom) sweeping magnetic field with varying pump polarization at time delays 13 ps (a-c) and 1.2ns (d-f). a,d) and b,e) present two separate magnetic field cycles, both using $\sigma^+$ -polarized pump, while c-f) are with linear-polarized pump. All data shown in (a-f) were acquired with a pump fluence of 42 μJcm$^{-2}$. g-l CD color maps ($\sigma^+$-$\sigma^-$) for the forward and reverse sweeping magnetic field with varying linear pump fluence at time delays 13 ps (g-i) and 1.2ns (j-l). Signal fluctuations are stochastic while critical fields remain similar across repeated scans, consistent with the instability of the non-polarized ν=-4/3 phase. Signal appears independent of whether linear vs circular polarized near-IR pump (~1220 nm) is used.

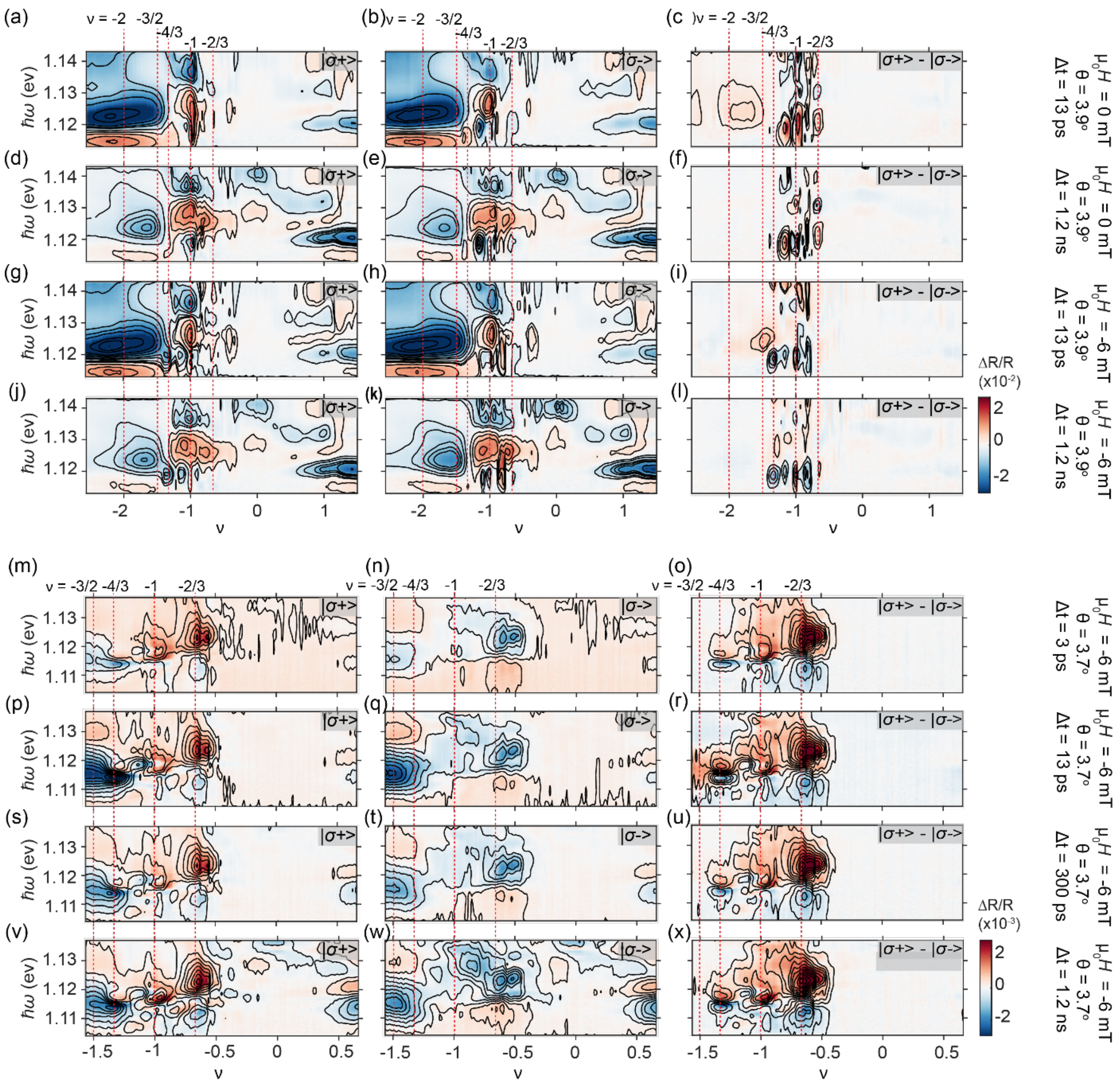

**Extended Data Fig 3. Transient circular dichroism (CD) gate maps at all time delays for D1 (θ=3.9°).** Spectrally resolved transient reflectance maps as a function of moiré filling factor (ν). The left and middle columns are ΔR/R spectra obtained with probe polarizations of $\sigma^+$ and $\sigma^-$, respectively, while the right column presents the difference spectra between the two probe polarizations ($\sigma^+$-$\sigma^-$) as a proxy to CD signal. For D1 (θ = 3.9°, a-f), data were taken at Δt = 13 ps and 1.2 ns at magnetic fields 0 mT and -6 mT. The complete gatemap is shown, covering a larger doping range shown in the main text. D2 (θ = 3.7°, m-x), data were taken at Δt = 3 ps, 13 ps, 300 ps and 1.2 ns at magnetic fields -6 mT. All data taken at $D_{eff}$ = 0 V/nm and 1.6 K.

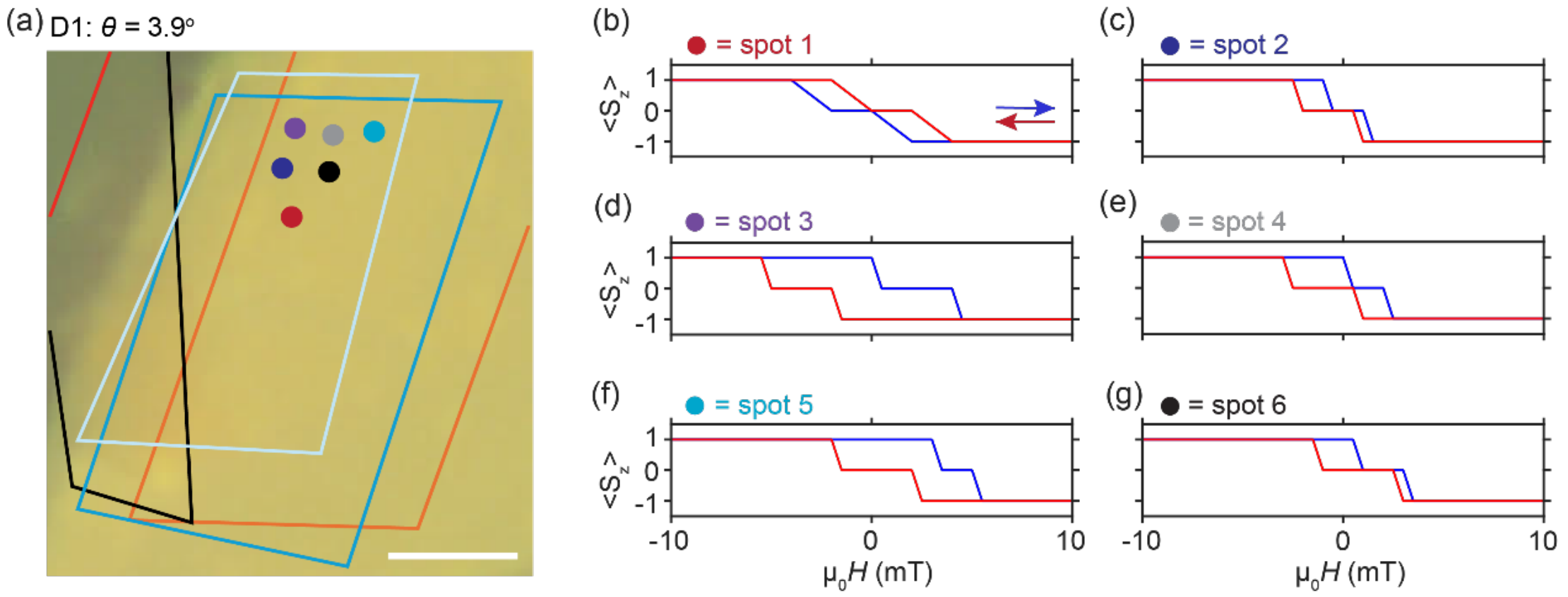

**Extended Data Fig 4**. **Sample position dependence of magnetic behavior at ν =-4/3 in D1 (θ =3.9° ).** This measurement was carried out following an additional warm-up and cool-down cycle after the measurement in Fig. 2a-f. (a) Zoom-In of optical image of D1 from Extended Data Fig 1 with measured sample positions on moiré bilayer region indicated in colored dots. Light and dark blue lines demarcate top and bottom monolayer $MoTe_2$ layers, respectively. Orange lines outline the top gate, black outlines the contact, and red outlines the bottom gate. White scale bar is 2.5μm. (b-g) Magnetic field dependence of the spin-valley polarization $\langle S_z \rangle = \frac{\Delta f_{AP}^{+} - \Delta f_{AP}^{-}}{\Delta f_{AP}^{+} + \Delta f_{AP}^{-}}$ sweeping in forward (blue) and reverse (red) directions at different sample positions. The selected spots indicate the region where the state exhibiting Ising AFM behavior at a filling factor of $\nu = -4/3$ is stabilized. The width of the ferromagnetic hysteresis loops observed in $\sigma^+$ and $\sigma^-$ vary spot to spot, a result likely due to a combination of twist angle inhomogeneity, strain, and variable screening from a spatially inhomogeneous dielectric environment. The reproducibility of the AFM behavior with spatially sensitive hysteretic responses are consistent with varying robustness of the two $\nu = -\frac{2}{3}$ copies in the FTI interpretation. We note that all data shown here were taken in a later thermal cycle than the main text figures. All data shown was taken at time delay Δt = 1.2 ns and $T$ = 1.6K.

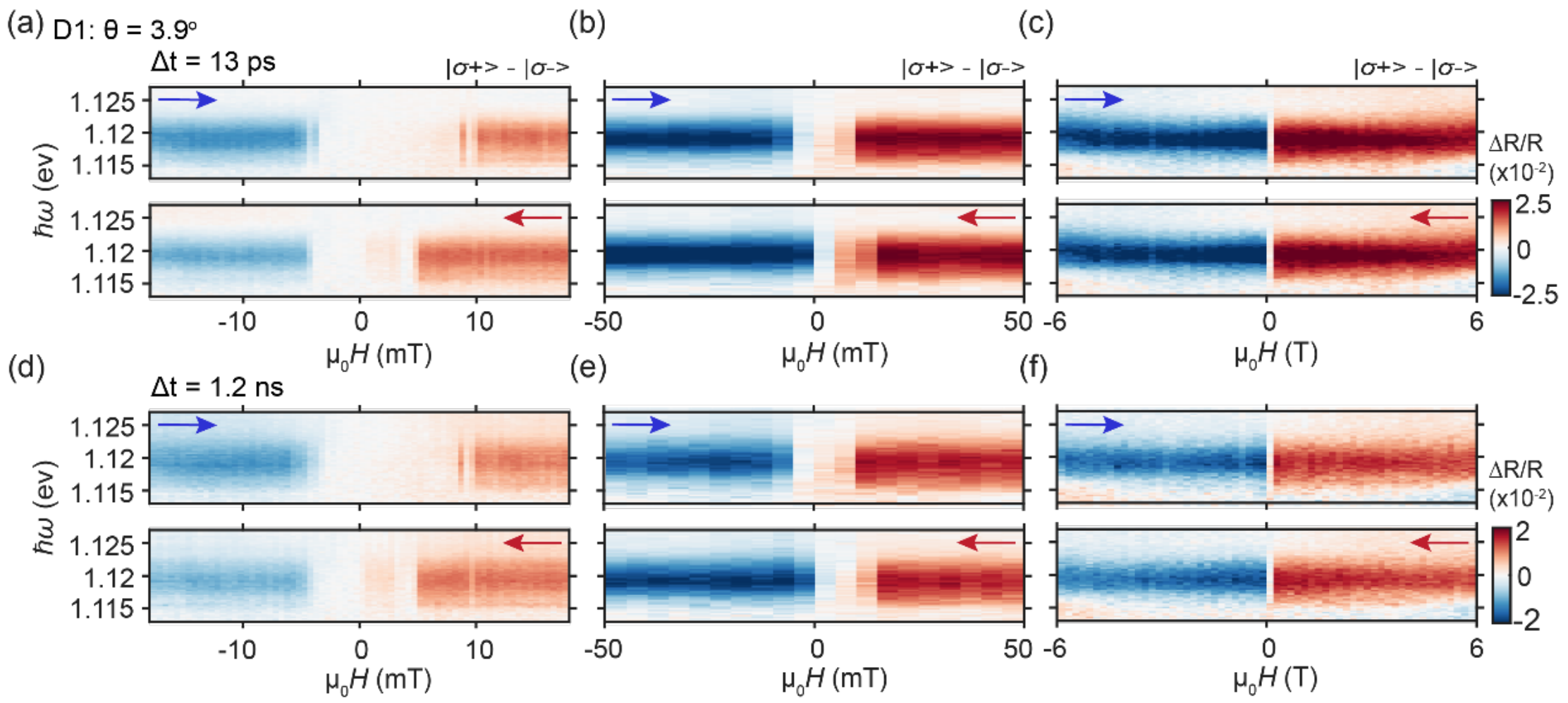

**Extended Data Fig 5. Mapping the high-field phase diagram for $\nu = -\frac{4}{3}$ in D1 (θ = 3.9°).** High field dependence of $\nu = -4/3$ at delay times a-c 13 ps and d-f 1.2 ns. The three columns present the difference spectra between the two probe polarizations $\sigma^+ - \sigma^-$ as a proxy to CD over a different magnetic field range with different step sizes. The first column (a,d) shows forward and reverse sweeps over $\pm$ 18 mT with a step size of 0.5 mT. Barring spin fluctuations, switching events occur $\leq$ |10| mT. The second column (b,e) shows magnetic field dependence over $\pm$ 50 mT with a larger step size of 5 mT, and the third (c,f) over $\pm$ 6T with a step size of 0.2T. Importantly, only a single switching event occurs in either direction $\leq \pm|150|$ mT. The valley polarized state is stable throughout the measured range, with a small Zeeman shift nearing $\pm$ 6T. This observation rules out a second phase transition to a Zeeman-induced band crossing state, where $\nu = -4/3$ resides in one valley and fills the second Chern band. Note that spin fluctuations, consistent with the fragile $\nu = -4/3$ unpolarized phase, lead to variations in intensity and the precise switching field across different scans. All data taken at 1.6K, D = V/nm.

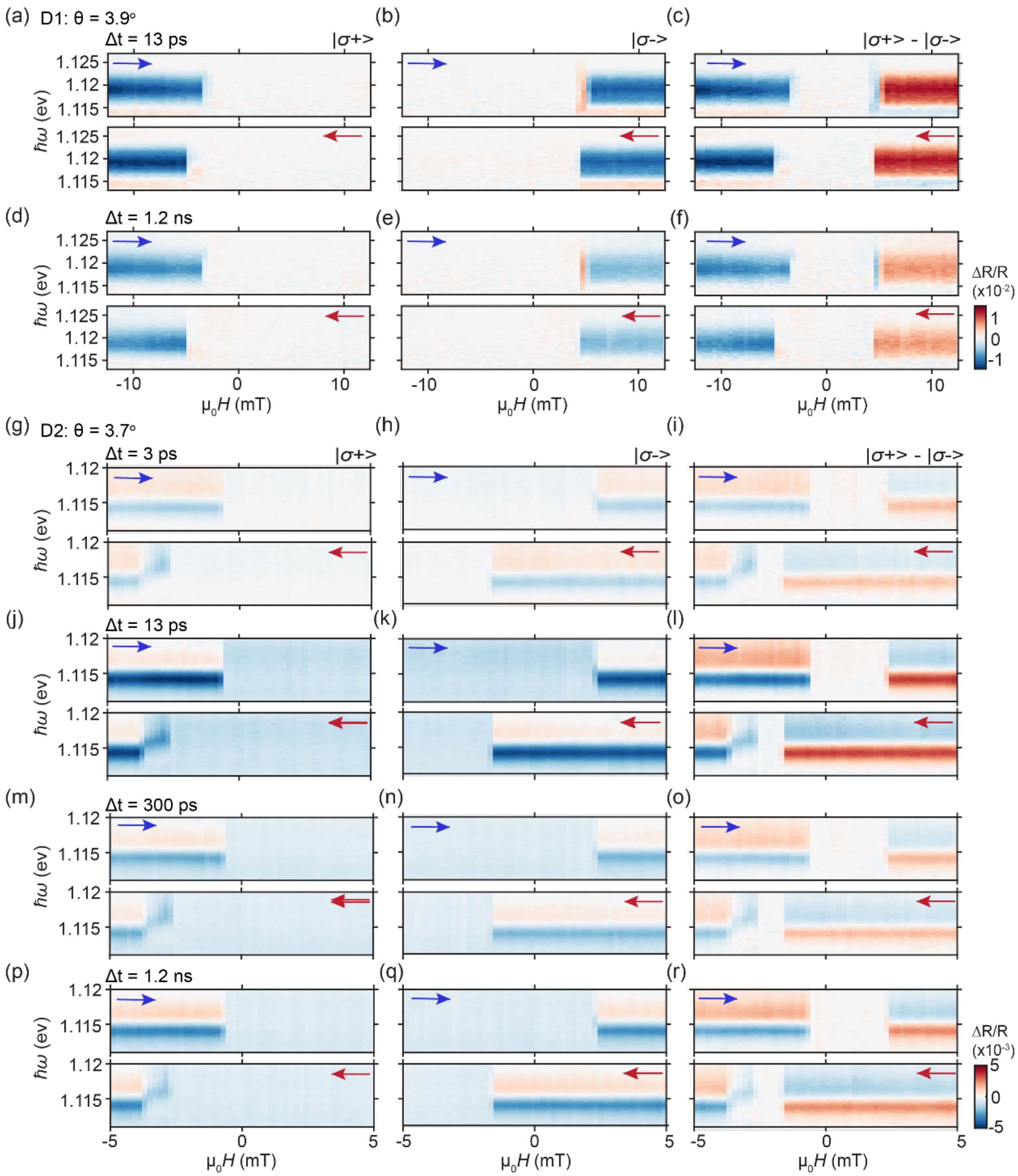

**Extended Data Fig 6**. **First order spin-flip transition occurs at same fields at different time delays.** Spectrally-resolved magnetic field dependence of AP feature in D1 (θ = 3.9deg) (a-f) and D2 (θ = 3.7deg) (g-r), respectively. The first column probes K with $\sigma^+$ -polarized light, the second column probes K' with $\sigma^-$ -polarized light, and the third shows the difference $\sigma^+$ - $\sigma^-$, serving as a proxy for CD. For each device, first-order spin flip transitions persist at all measured time delays with identical switching fields (<10mT), differing only in intensity due to melting/recovery dynamics. Note that both devices exhibit three regions which remain flat with respect to magnetic field, including two magnetic regions and one non-magnetic region near 0mT.

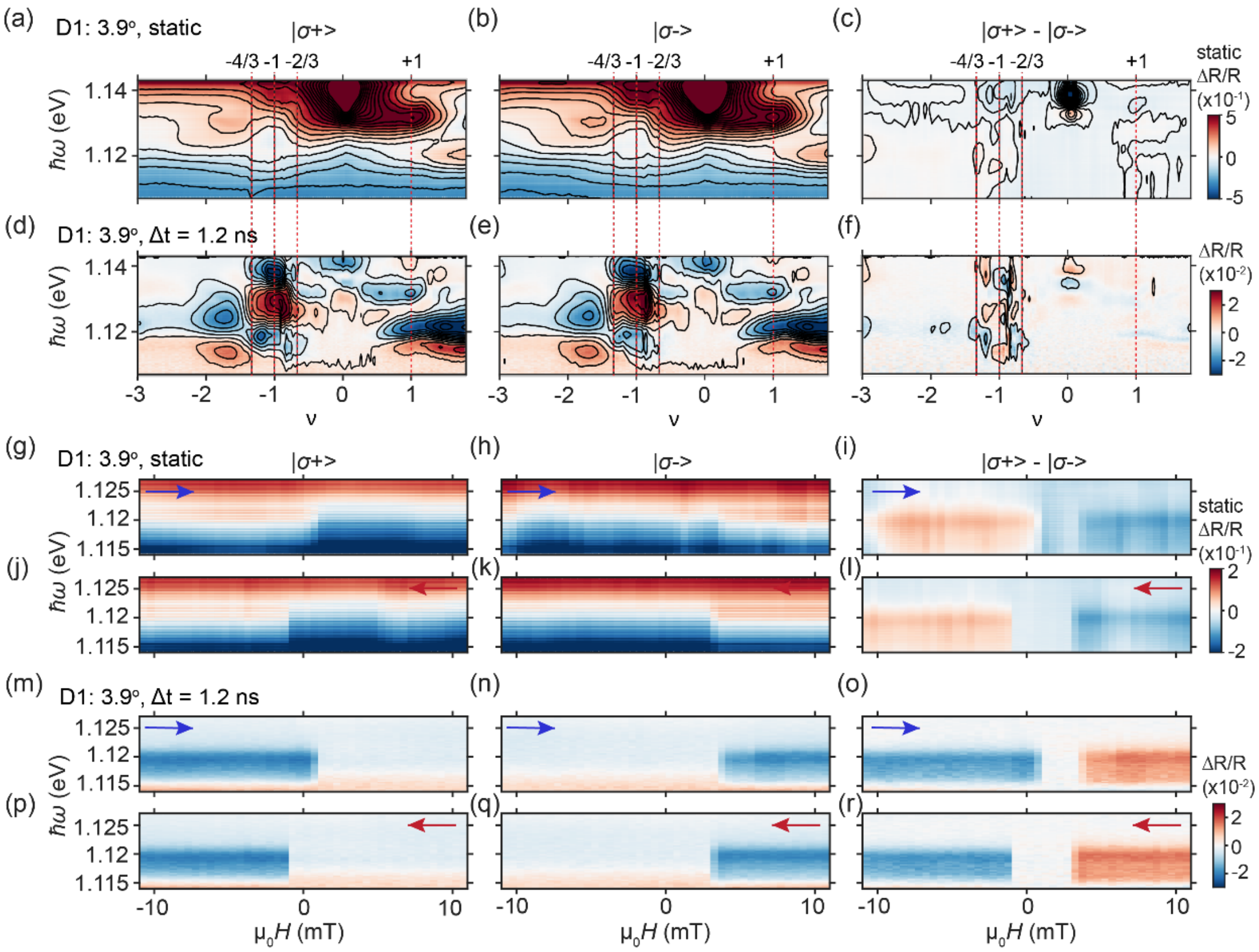

**Extended Data Fig. 7. Static versus transient CD spectra.** Static reflectance of sample D1 (θ = 3.9°) at external magnetic fields of $\mu_0H$ = -6 mT probed by (a) $\sigma^+$ and (b) $\sigma^-$ polarized light. In the static $\Delta R/R$ spectra, $\Delta R$ is given by $R_{on\ sample} - R_{off\ sample}$ and is normalized to background $R_{off\ sample}$, which represents the probe spectrum on a region of the device with all layers except for bilayer $MoTe_2$. Panel (c) presents the difference spectra between the two probe polarizations, ($\sigma^+$-$\sigma^-$), as proxy to CD spectra. There is no pump pulse in the static spectra. (d)-(f) is the transient version of (a)-(c) at a time delay of 1.2ns, where higher resolution of states is obtained under pump modulation. In the transient $\Delta R/R$ spectra, $\Delta R$ is given by $R_{pump\ on} - R_{pump\ off}$ and is normalized to background $R_{pump\ off}$. The red-dashed lines mark filling factors of ν = -4/3, -1, -2/3, +1. All (g)-(k) shows the static reflectance response of the ν =-4/3 state to out-of-plane magnetic field in forward and reverse sweeping directions, where (i)-(l) shows the difference ($\sigma^+$-$\sigma^-$). (m-r) shows the transient version of (g)-(l). The transient spectra exhibit near identical AFM behavior compared with the static measurements, including the same critical fields where switching between the nonpolarized and partially polarized phases are observed. This figure demonstrates the improved sensitivity of the pump-probe approach to detecting correlated phases and signatures of magnetization which are intrinsic to the sample in the absence of pump modulation. All spectra obtained at sample stage temperature of T = 1.6 K.

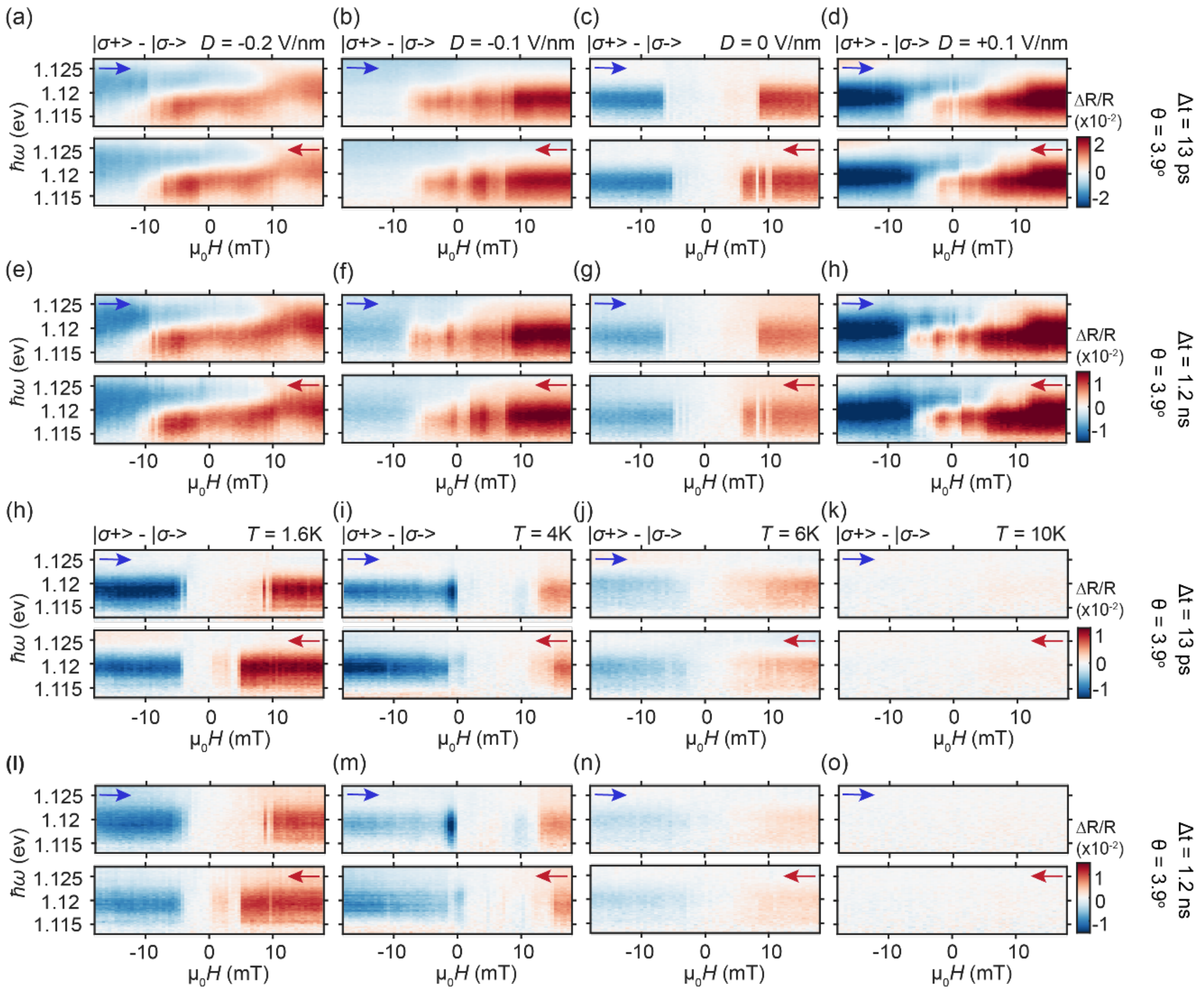

**Extended Data Fig 8**. **Robustness of ν =-4/3 in D1 (θ =3.9°) to displacement-field and temperature.** (a-h) magnetic field dependence of $\sigma^+$ - $\sigma^-$ , signal sweeping in forward and reverse direction as a function of displacement field (i-p) and temperature at time delays of 13 ps and 1.2 ns.

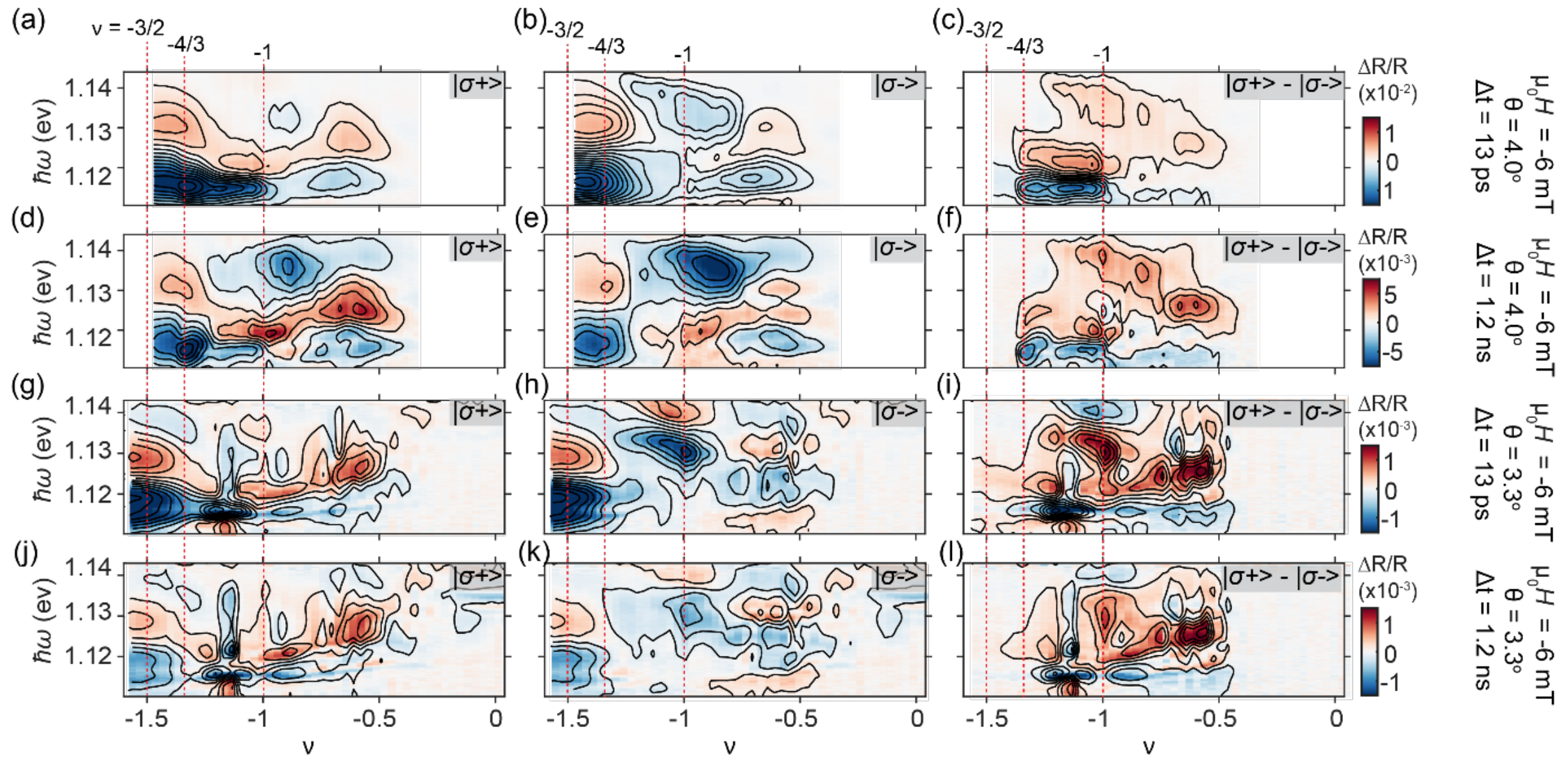

**Extended Data Fig 9. Gate maps and magnetic field dependence of other twist angles (D3 θ =4.0°) and (D4 θ =3.3°)** Spectrally-resolved gate maps as a function of filling factor on the hole-doped side. The first column shows $\sigma^+$-polarized light, the second column shows signal probed by $\sigma^-$-polarized light, and the third column shows the difference (CD) for (a-f) D3 and (g-l) D4. Red dotted lines indicate states at $\nu$ =-3/2, -4/3, and -1.

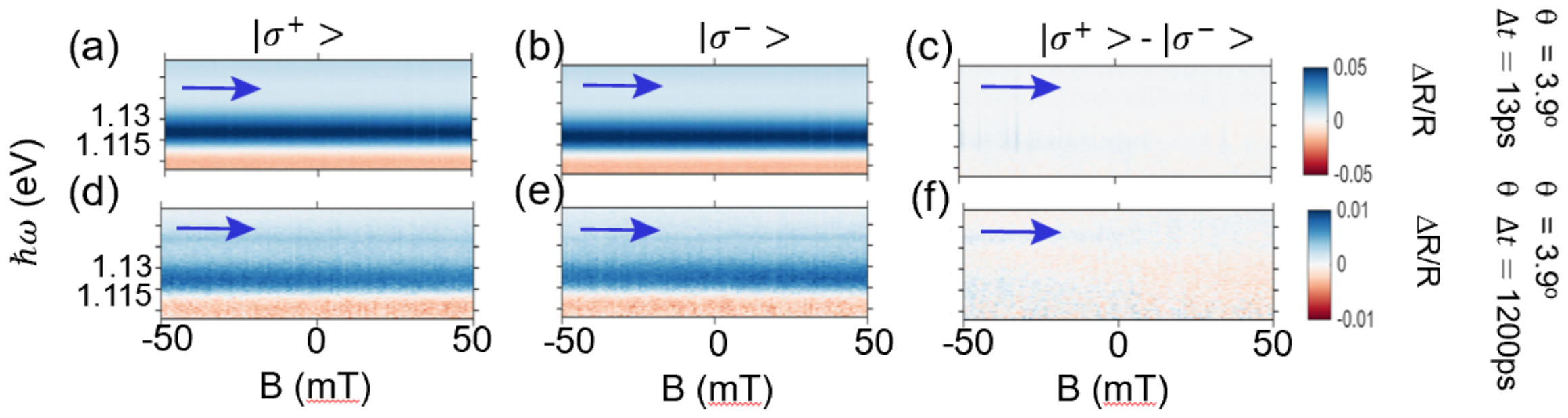

**Extended Data Fig 10. Magnetic field dependence of $\nu = -2$ in D1 (θ = 3.9deg), -50mT → +50 mT (raw data, no bg correction).** The first column shows signal probed by $\sigma^+$-polarized light, the second column shows signal probed by $\sigma^-$-polarized light, and the third column shows the difference (CD). The Δ R/R signal is completely flat in the measured range ± 50mT, consistent with a robust non-polarized state at a filling factor of ν =-2.

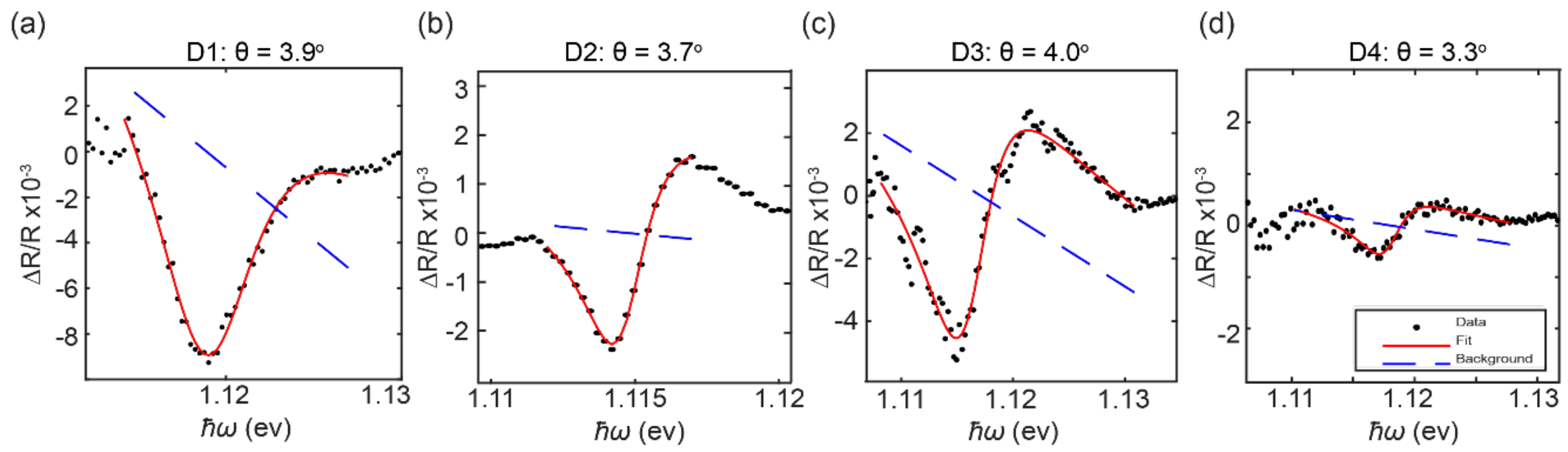

**Extended Data Fig. 11. Fitting Δ R/R for AP signal in the four measured devices.** Sample fits of spectral cuts where there is RMCD signal at a time delay of 1200 ps. See methods for details.

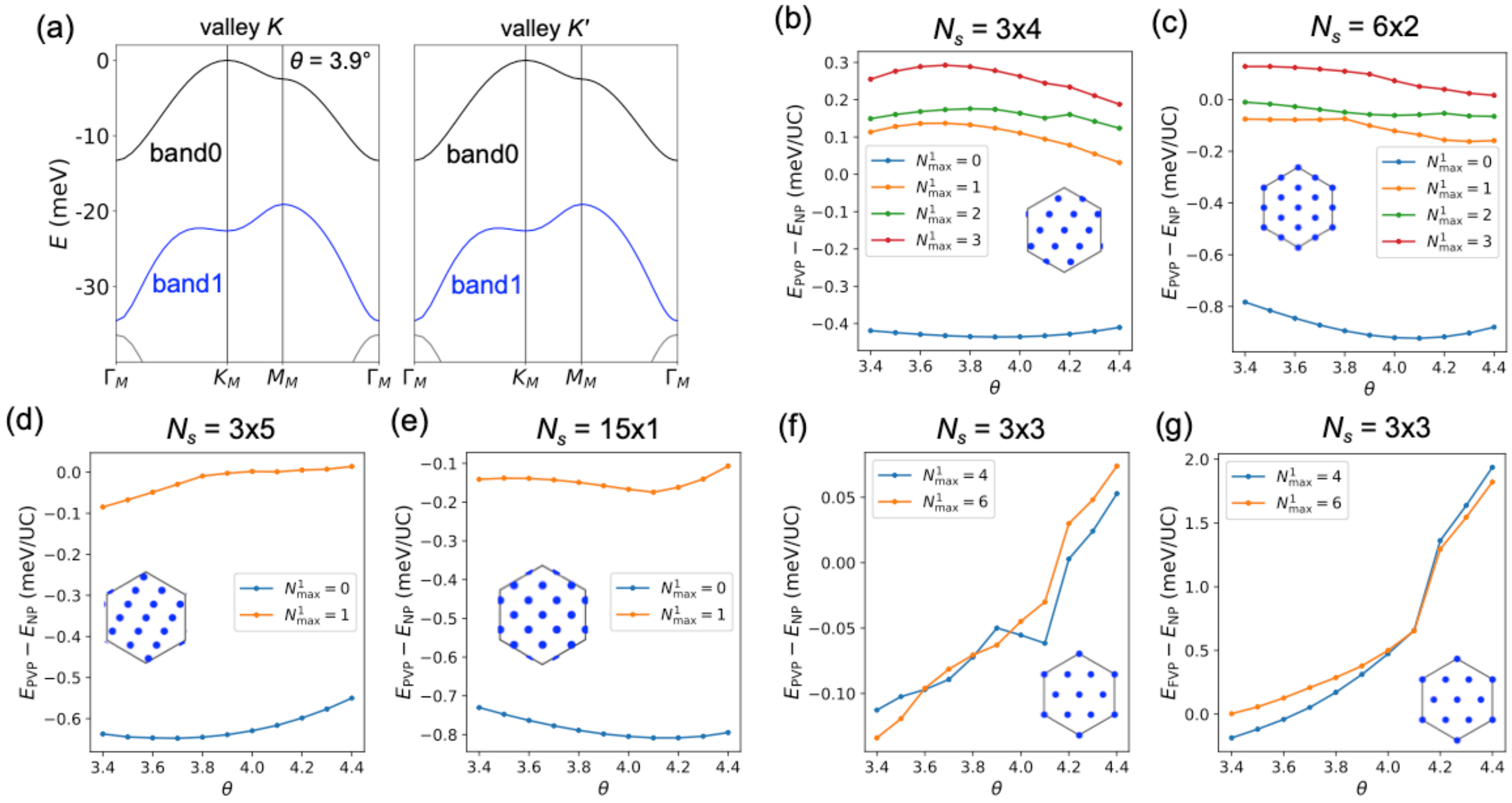

**Extended Data Fig 12**. **Band-mixing exact diagonalization calculations of valley polarization** at $\nu = -4/3$**.** a) Band structure of $tMoTe_2$ at $\theta = 3.9^{\circ}$ for the two valleys. The two lowest moiré valence bands are labelled as band0 and band1. b)-f) Energy difference per unit cell between the partially valley-polarized (PVP) sector ($\nu_+ = -1$ and $\nu_- = -1/3$) and the non-polarized (NP) sector ($\nu_+ = \nu_- = -2/3$) for different twist angles $\theta$ and band-mixing parameters $N^1_{max}$. The panels refer to different system sizes $N_s$. The mBZ momentum mesh is shown as inset. g) Energy difference per unit cell between the fully valleypolarized (FVP) sector ($\nu_+ = -4/3$ and $\nu_- = 0$) and the NP sector for $N_s = 3 \times 3$.

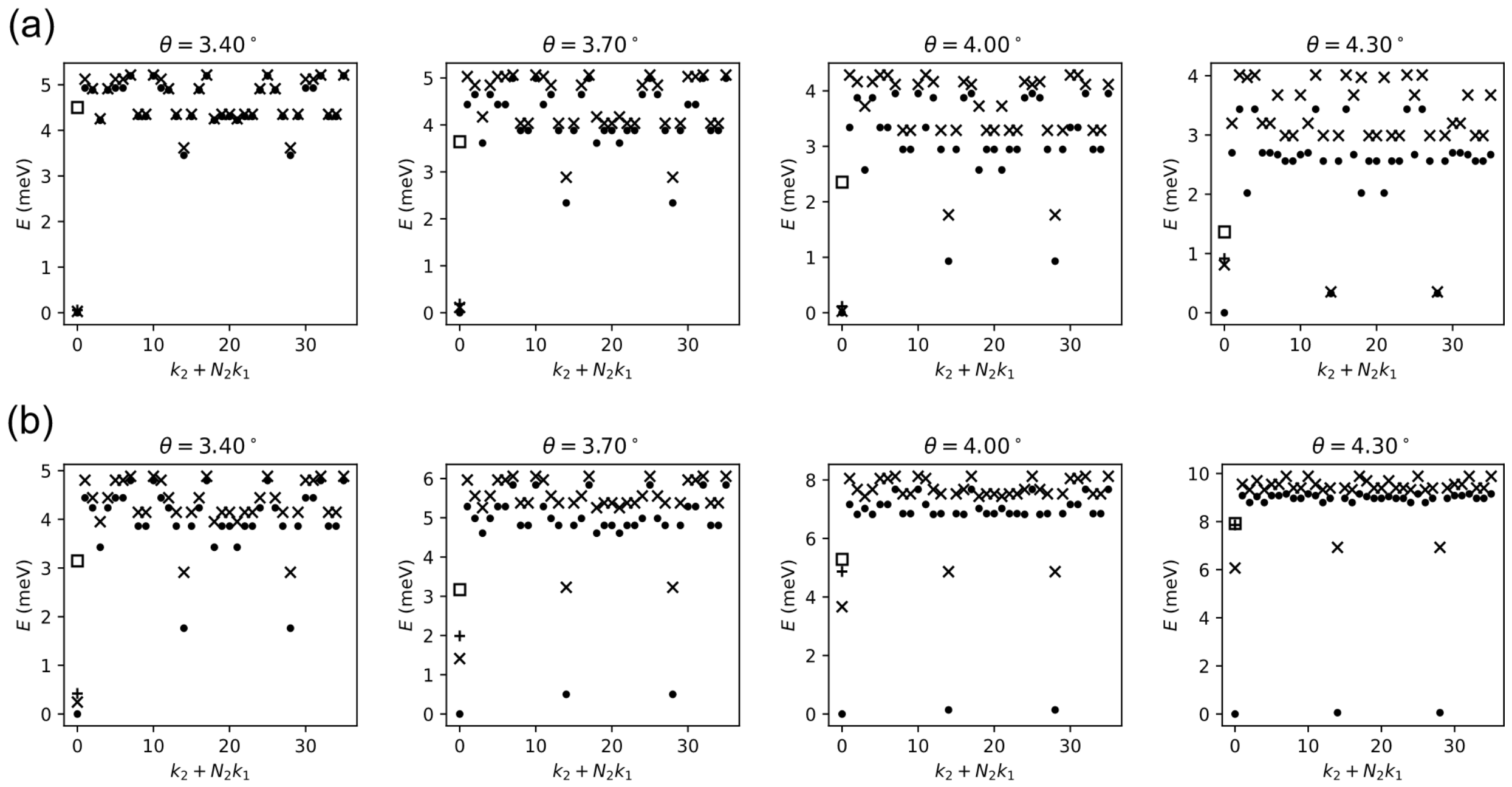

**Extended Data Fig 13**. **Single-band exact diagonalization calculations of fully valley-polarized states for** $N_s = 6{\times}6$**.** a) Many-body spectra for $\nu_+ = -2/3, \nu_- = 0$. A $\nu = -2/3$ FCI is characterized by a topological three-fold degenerate gapped ground state at $(k_1,k_2) = (0,0)$. For a given many-body momentum, the symbols cycle between $\cdot, \times, +, \square$ to help distinguish nearly-degenerate states. b) Same as a) except for $\nu_+ = -1/3, \nu_- = 0$. A $\nu = -1/3$ FCI is characterized by a topological three-fold degenerate gapped ground state at $(k_1,k_2) = (0,0)$. A $\nu = -1/3$ $K_M$-CDW is characterized by three degenerate gapped ground states at momenta $(k_1,k_2) = (0,0),(2,2),(4,4)$.

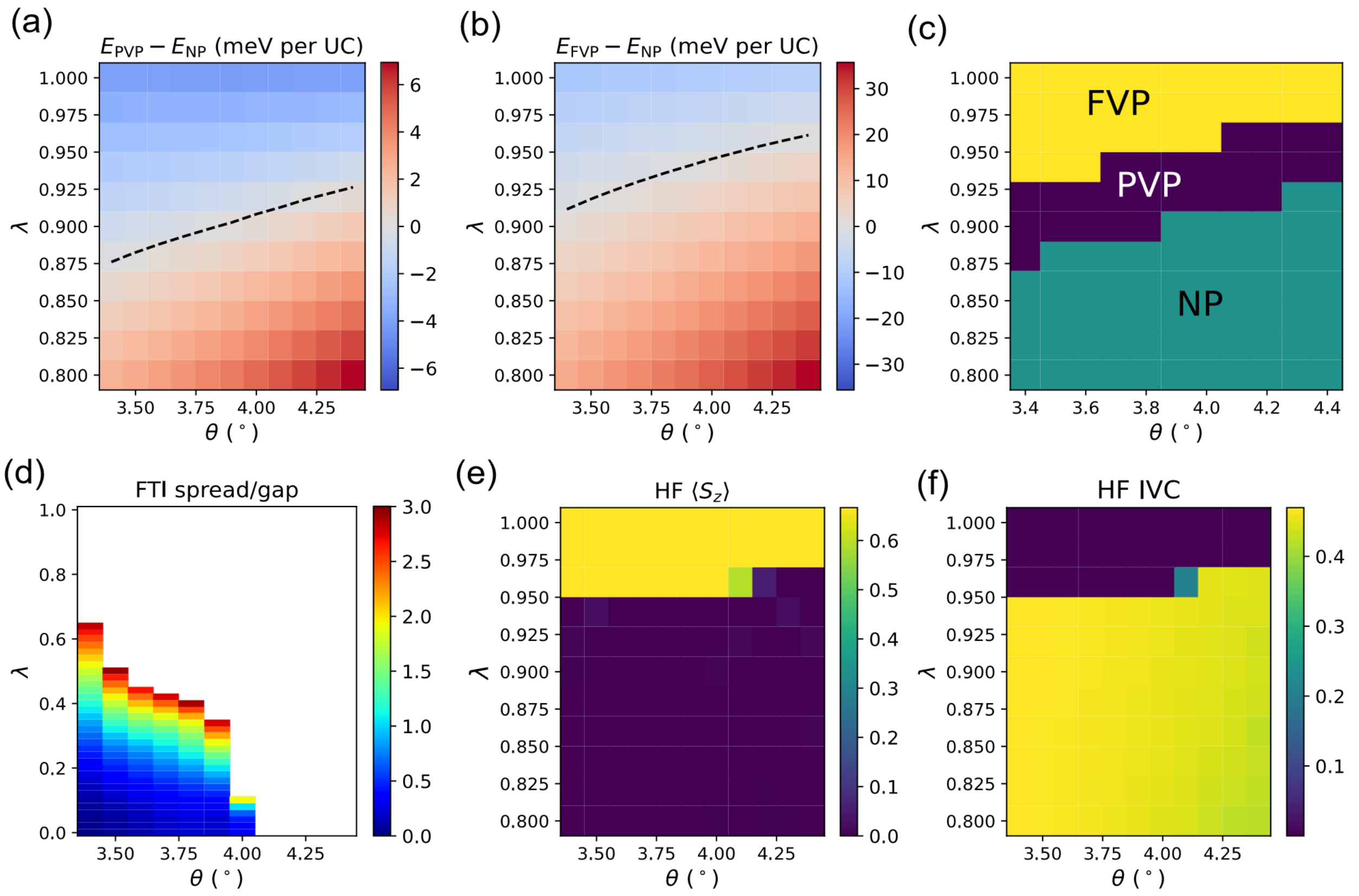

**Extended Data Fig 14**. **Calculations at** $\nu = -4/3$ **for** $N_s = 6{\times}6$**.** a) Comparison of PVP (see Sec. 8.3.3) and NP (see Sec. 8.3.2) energies, as a function of intervalley parameter $\lambda$ and twist angle $\theta$. b) Comparison of FVP (see Sec. 8.3.5) and NP energies. c) Phase diagram showing the favored polarization sector in the ED calculations. d) FTI spread/gap ratio for the NP sector. e) Average valley polarization $\langle S_z \rangle$ in the HF calculation. f) Magnitude of IVC in the HF calculation.

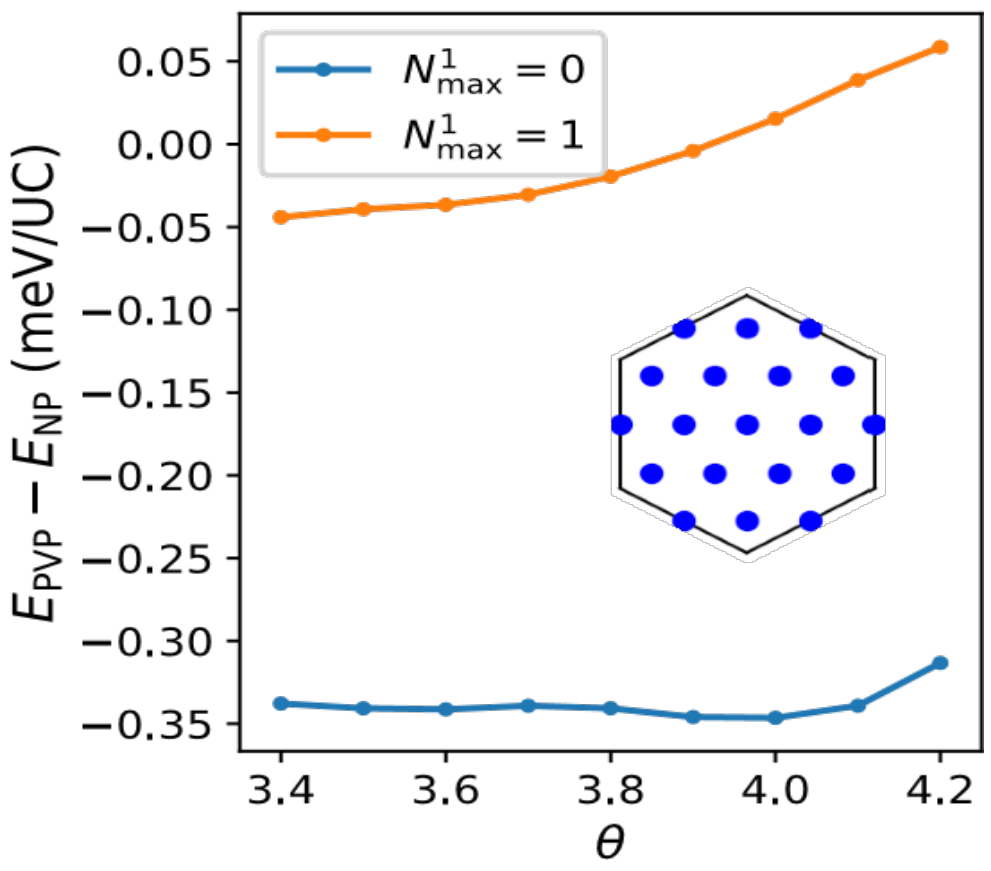

**Extended Data Fig 15**. **Band-mixing exact diagonalization calculations of valley polarization at** $\nu = -3/2$**.** Energy difference per unit cell between the partially valley-polarized (PVP) sector ($\nu_+ = -1$ and $\nu_- = -1/2$) and the non-polarized (NP) sector ($\nu_+ = \nu_- = -3/4$) for different twist angles $\theta$ and band-mixing parameters $N^1_{max}$. The mBZ momentum mesh is shown as inset.